\documentclass{article}

\usepackage{arxiv}

\usepackage[utf8]{inputenc} 
\usepackage[T1]{fontenc}  
\usepackage{hyperref}    
\usepackage{url}      
\usepackage{booktabs}    
\usepackage{amsfonts}    
\usepackage{nicefrac}    
\usepackage{microtype}   
\usepackage{lipsum}		
\usepackage{graphicx}
\usepackage{natbib}
\usepackage{doi}
\usepackage{amsmath}
\usepackage{mathtools}
\usepackage{enumitem}

\usepackage{cleveref}

\usepackage{tikz}
\usepackage{amsthm, amssymb}
\usepackage{graphicx}
\usepackage{mathabx}

\usepackage{makecell}
\usepackage{algorithm}
\usepackage{algpseudocode}
\usepackage[utf8]{inputenc}
\usepackage{array}

\usepackage{subcaption}
\usepackage[export]{adjustbox}

\newtheoremstyle{indented}{3pt}{3pt}{\addtolength{\leftskip}{2.5em}}{}{\bfseries}{.}{.5em}{}
\theoremstyle{indented}
\newtheorem{theorem}{Theorem}
\newtheorem{property}{Property}

\title{Causal Inference for Latent Outcomes\\Learned with Factor Models}

\author{ 
  \href{https://orcid.org/0000-0002-1411-8750}{\includegraphics[scale=0.06]{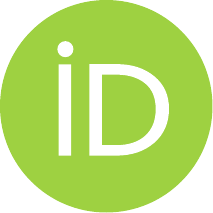}\hspace{1mm}Jenna M. Landy} \\
	Harvard University\\
	\texttt{jlandy@g.harvard.edu} \\  
    \And
	\href{https://orcid.org/0009-0006-5228-2953}{\includegraphics[scale=0.06]{orcid.pdf}\hspace{1mm}Dafne Zorzetto} \\
	Brown University\\
  \texttt{dafne\_zorzetto@brown.edu}\\
    \And
	\href{https://orcid.org/0000-0003-0639-5341}{\includegraphics[scale=0.06]{orcid.pdf}\hspace{1mm}Roberta De Vito} \\
	 Sapienza University of Rome\\
    Brown University\\
  \texttt{roberta.devito@uniroma1.it}\\
    \And
	\href{https://orcid.org/0000-0002-8783-5961}{\includegraphics[scale=0.06]{orcid.pdf}\hspace{1mm}Giovanni Parmigiani} \\
	Dana Farber Cancer Institute\\
  Harvard University\\
  \texttt{gp@jimmy.harvard.edu}
}

\renewcommand{\shorttitle}{Causal Inference for Latent Outcomes Learned with Factor Models}

\hypersetup{
pdftitle={Causal Inference for Latent Outcomes Learned with Factor Models},
pdfsubject={q-bio.NC, q-bio.QM},
pdfauthor={Jenna Landy, Giovanni Parmigiani},
pdfkeywords={Causal Inference, Latent Outcomes, Latent Factor Models, nonnegative Matrix Factorization, Mutational Signatures Analysis},
}

\begin{document}
\maketitle

\begin{abstract}
  In many fields---including genomics, epidemiology, natural language processing, social and behavioral sciences, and economics---it is increasingly important to address causal questions in the context of factor models or representation learning. In this work, we investigate causal effects on \textit{latent outcomes} derived from high-dimensional observed data using nonnegative matrix factorization. To the best of our knowledge, this is the first study to formally address causal inference in this setting. A central challenge is that estimating a latent factor model can cause an individual's learned latent outcome to depend on other individuals' treatments, thereby violating the standard causal inference assumption of no interference. We formalize this issue as \textit{learning-induced interference} and distinguish it from interference present in a data-generating process. To address this, we propose a novel, intuitive, and theoretically grounded algorithm to estimate causal effects on latent outcomes while mitigating learning-induced interference and improving estimation efficiency. We establish theoretical guarantees for the consistency of our estimator and demonstrate its practical utility through simulation studies and an application to cancer mutational signature analysis. All baseline and proposed methods are available in our open-source R package, {\tt causalLFO}.
\end{abstract}


\section{Introduction}

Causal inference aims to quantify the effect of a specific cause or exposure on an outcome of interest. Traditionally, such outcomes are low-dimensional and directly observable, such as disease onset, number of citations, or sale price. However, the increasing amount of high-dimensional data poses a new challenge in causal inference: outcomes are becoming more complex and multivariate, such as mutation counts in tumor genomes, word frequencies in documents, or daily stock prices. These complex measurements can be simplified and made more interpretable through factor models which consider each high-dimensional observation as a function of underlying unobserved latent representations. Examples of such latent representations include mutational processes or signatures in a tumor genome, topics used in a document, or stock market sectors. In this paper, we focus on nonnegative matrix factorization (NMF), a factor modeling technique for decomposing nonnegative data into nonnegative factors and weights, yielding a highly interpretable parts-based representation \citep{lee1999learning}.

Causal questions in the context of latent factors and representation learning are of growing interest, as evidenced by the growing body of literature over the past decade. Previous works have incorporated latent variables as treatments \citep{fong2016discovery, feder2022causal, egami2022make, knox2022testing, vanderweele_constructed_2022}, matching variables \citep{roberts_adjusting_2020}, covariates or confounders \citep{lee_estimation_2018,keith-etal-2020-text, feder2022causal, knox2022testing}, for imputation \citep{athey2021matrix,vega2024spatio}, and to account for correlation structures between multidimensional outcomes \citep{zorzetto2025multivariate}.

However, the literature on \textit{latent outcomes} is limited. In social and behavioral sciences, latent factors are used as proxies for outcomes that are conceptual abstractions like ``democracy'' or ``perception''. These applications often employ a two-step procedure that first estimates latent outcomes using a latent outcome model, often factor analysis, and then applies traditional causal inference methods to estimate the treatment effect \citep{knox2022testing}. In this paper, we refer to this procedure as the \textit{All Data algorithm}. Similar approaches have been adopted in natural language processing (NLP), where latent outcomes may be derived from topic models or language models \citep{feder2022causal, egami2022make}. To the best of our knowledge, ours is the first work to formally address causal inference on latent outcomes derived from NMF.

A key methodological challenge in this setting is what we name \textit{learning-induced interference}. As noted in prior NLP work \citep{egami2022make,feder2022causal}, when latent outcome models are trained on the full dataset, the resulting representations for an individual may depend on the treatment assignments of others, violating the no-interference assumption central to most causal inference frameworks. A second challenge encountered in this work is high variability in causal estimates, especially in highly heterogeneous settings like mutational signatures analysis.

To address these challenges, we make two main contributions. First, we formalize the concept of learning-induced interference and distinguish it from standard interference in a data-generating process. To this scope, we provide a motivating example illustrating the degree of impact that learning-induced interference can have on learned latent outcomes and causal estimates if current approaches are used. Second, we develop the \textit{Impute and Stabilize algorithm} to estimate causal average treatment effects on latent outcomes.  This algorithm reduces the magnitude of learning-induced interference by fitting the factor model only on untreated individuals, thereby stabilizing the input to NMF and making the factor model less sensitive to treatments. Efficiency is gained by imputing unobserved potential outcomes to allow for a larger sample size for factorization and for paired contrasts that account for sample-to-sample variation. We provide theoretical guarantees of consistency under a set of additional assumptions. Additionally, we develop a bootstrap wrapper and align factor models across bootstrap repetitions to quantify uncertainty and make significance decisions.

We evaluate the proposed approach through simulation studies. We show that the Impute and Stabilize algorithm has unbiased average treatment effect estimates, even when one or more baseline approaches are biased. Further, Impute and Stabilize shows a significant efficiency improvement, particularly for factors whose weights have outliers, and efficiency comparable to baselines otherwise. Borrowing from literature on interference in vaccine trials, we quantify indirect effects in simulation studies and show that our method reduces the degree of learning-induced interference by a factor of at least two compared to baseline approaches. Finally, we apply all algorithms in the context of cancer mutational signatures analysis to estimate the effect of a germline BRCA mutation on the contributions of signatures in early-onset breast adenocarcinoma. We provide an open-source R software package, {\tt causalLFO} implementing all algorithms discussed, available on GitHub at \href{https://github.com/jennalandy/causalLFO}{jennalandy/causalLFO}.

\section{Background and definitions}

\subsection{Notation and causal estimand}
\label{sec:causal-notation}

Let $i$ index subjects $1,\dots,N$ and $d$ index variables $1,\dots,D$. The relationships among all variables are represented in the causal DAG of Figure \ref{fig:dag0}. We assume binary treatments assigned through a completely randomized experiment with no covariates, such that $\mathbf{T} \in \{0,1\}^N$ is our randomized binary treatment vector, or treatment program, indicating whether each subject $i$ is treated ($T_i = 1$) or untreated ($T_i = 0$). The matrix $\mathbf Y$ contains observed post-treatment data, with each column $Y_i$ representing a $D$-dimensional data vector for subject $i$. For any set of subject indices $I$, $\mathbf Y_I$ denotes a subset of the data matrix corresponding to columns with indices $i \in I$. Under the potential outcomes framework, $Y_i(\mathbf t)$ denotes the data vector for subject $i$ in the counterfactual world where $\mathbf{T}$ is set to the realized treatment vector $\mathbf{t} = [t_1, ..., t_N]$.

We define $\mathbf{L}$ as a matrix of \textit{latent outcomes} and index latent dimensions $k = 1,\dots, K$, where each column $L_i$ represents a $K$-dimensional latent outcome for individual $i$. We assume that $L$ is an intermediate between $T$ and $Y$ so that the treatment affects observed data exclusively through the latent outcome $L$. This setup differs from standard causal mediation analysis \citep{robins1992identifiability} and principal stratification \citep{frangakis2002principal}, as the latent outcome $L$ is not merely a mediator or post-treatment variable, but rather the primary outcome of interest. Our goal is to estimate the causal effect of $T$ on $L$, not on $Y$. As a post-treatment variable, $L$ has potential outcomes $L_i(\mathbf{t})$, the latent outcome vector for subject $i$ in the counterfactual world where $\mathbf{T}$ is set to $\mathbf{t}$. However, in the case of latent outcomes, we do not directly observe any values of $L(\mathbf{t})$.

We adopt the standard causal inference assumption of no interference in the data generating model, meaning an individual's outcome depends only on their own treatment: $L_i(\mathbf t) = L_i(t_i)$ \citep{cox1958interpretation,rubin1974estimating, rubin1980randomization}. This assumption on the data generating model can be extended to the observed data to assume $Y_i(\mathbf{t}) = Y_i(t_i)$. Then, the causal effect of interest, the average treatment effect (ATE) on $L$, is the difference in expected values of latent potential outcomes under treatment versus no treatment:
\begin{align*}
  \boldsymbol{\psi}_{L} &= \mathbb E[L(1)] - \mathbb E[L(0)], \quad\quad \psi_{L_k} =\mathbb E[L_k(1)] - \mathbb E[L_k(0)]
\end{align*}
where $\boldsymbol{\psi}_{L}$ is a K-dimensional vector, matching the dimensionality of $L$.

We also assume standard casual inference conditions of consistency, positivity, and exchangeability due to complete randomization which enable identification: $\mathbb E[ Y(t)] = \mathbb E[ Y|T = t]$ and $\mathbb E[ L(t)] = \mathbb E[ L|T = t]$ \citep{rubin1974estimating}.

\usetikzlibrary{arrows.meta, positioning, calc}

\begin{figure}[t]
    \centering
    \begin{tikzpicture}[
        roundnode/.style={circle, draw=black, thick, minimum size=1.4cm, font=\large, fill=white},
        roundnodesmall/.style={circle, draw=black, thick, minimum size=0.1cm, font=\large, fill=white},
        arrow/.style={-Stealth, thick},
        estimated/.style={circle, draw=black, fill=gray!20, thick, minimum size=1.4cm, font=\large},
        estimatedsmall/.style={circle, draw=black, fill=gray!20, thick, minimum size=0.1cm, font=\large},
        redarrow/.style={-Stealth, thick, red}
      ]
    
      \node[roundnode] (Ti) at (0,0) {$T_i$};
      \node[estimated] (Li) at (2,0) {$L_i$};
      \node[roundnode] (Yi) at (4,0) {$Y_i$};
    
      \node[roundnode] (Tj) at (0,-1.8) {$T_j$};
      \node[estimated] (Lj) at (2,-1.8) {$L_j$};
      \node[roundnode] (Yj) at (4,-1.8) {$Y_j$};
    
      \node[estimated] (lambda) at (2,1.8) {$\boldsymbol \lambda$};
      \node[roundnode] (Y) at (5.8,-0.9) {$\mathbf{Y}$};
    
      \draw[redarrow] (Ti) -- (Li);
      \draw[arrow] (Li) -- (Yi);
      \draw[redarrow] (Tj) -- (Lj);
      \draw[arrow] (Lj) -- (Yj);
    
      \draw[arrow] (lambda) -- (Yi);
      \draw[arrow] (lambda) -- (Yj);
    
      \node at ($(Yi)!0.4!(Y)$) [rotate=-30, anchor=south west, font=\Large] {$\in$};
      \node at ($(Yj)!0.4!(Y)$) [rotate=30, anchor=north west, font=\Large] {$\in$};
    
      \begin{scope}[xshift=10cm, yshift=-0.5cm]
      \node[rectangle, draw=black, thick, fill=white, inner sep=8pt] (leg) at (0,0) {
        \begin{tabular}{@{}m{1cm} m{3.5cm}@{}}
          \tikz{\draw[arrow] (0,0) -- (0.6,0);} & Causal effect \\
          \tikz{\draw[redarrow] (0,0) -- (0.6,0);} & Causal effect of interest \\
          \tikz{\node[estimatedsmall] at (0,0) {}; } & Unobservable \\
          \tikz{\node[roundnodesmall] at (0,0) {}; } & Observable \\
        \end{tabular}
      };
    \end{scope}
    
    \end{tikzpicture}
    \caption{\textbf{Causal DAG with ground truth} (unobservable) latent factors $\boldsymbol \lambda$ and latent outcomes $L$. Assuming no interference in the data generating process, there is no path from $T_i$ to $L_j$ or from $T_j$ to $L_i$ for each $i \ne j$.}
    \label{fig:dag0}
\end{figure}
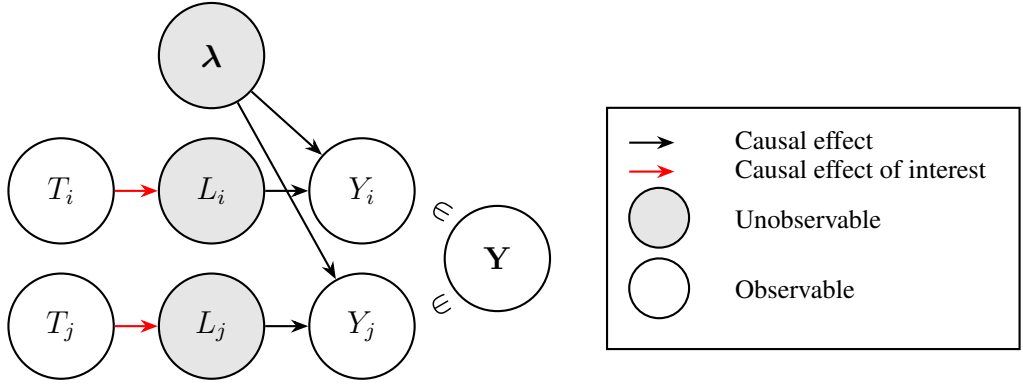

\subsection{NMF and NMF-learned outcomes}
We now place the latent outcome $L$ and observed data $Y$ in the context of NMF and NMF-learned outcomes. NMF is a popular method in representation learning for interpretable parts-based representation of nonnegative data, for example in mutational signatures analysis, document topic modeling, image processing, and financial portfolio analysis \citep{wang2012nonnegative}.

NMF decomposes the full observed data matrix $\mathbf Y$ into two lower-rank matrices, factors $\boldsymbol \lambda \in \mathbb{R}_{\ge 0}^{D \times K}$ and contributions $\mathbf L \in \mathbb{R}_{\ge 0}^{K \times N}$, such that $\mathbf Y \approx \boldsymbol \lambda \mathbf L$ where the number of factors $K\!\!\ll\!\!N, D$ \citep{paatero1994positive, lee1999learning}. We assume Poisson-generated data as follows:
\begin{align*}
  Y_{di} \sim \text{Poisson}\left(\sum_{k = 1}^K \lambda_{dk}L_{ki}\right)
\end{align*}
where $\lambda_{dk}$ is the single element in the factor matrix $\boldsymbol \lambda$, such that each column $\lambda_k$ represents a single factor. The latent outcomes of interest are the columns of $\mathbf L$, commonly referred to as the factor weights, loadings, or contributions matrix.

NMF parameters are traditionally estimated through gradient descent to minimize a reconstruction error with multiplicative weights to maintain non-negativity. Maximizing the Poisson likelihood with the Expectation-Maximization (EM) algorithm \citep{dempster1977maximum} is equivalent to minimizing Kullback–Leibler (KL) divergence
\citep{kullback1951information} with this gradient descent approach. For all instances of NMF, we use the {\tt NMF} R software package with the {\tt "brunet"} algorithm option to minimize KL-divergence \citep{nmf_R}.

If the factor matrix $\boldsymbol \lambda$ is fixed, a nonnegative linear model (NNLM) can be used to estimate $\mathbf L$ from $\mathbf Y$ (for a Gaussian likelihood, commonly referred to as nonnegative least squares, or NNLS). For all instances of NNLM, we use our own implementation of gradient descent to minimize KL-divergence with a fixed factor matrix, using the standard multiplicative updates from \citet{lee1999learning}.

NMF-learned outcomes often have concrete interpretations grounded in the application. For instance, in mutational signatures analysis, latent outcomes represent the number of mutations attributed to distinct mutational processes. In other genomics settings, they may reflect latent biomarkers or composite phenotypes derived from high-dimensional molecular data.

\subsection{Learning latent outcomes and learning-induced interference}

As an unobservable variable, our latent outcome of interest $L$ must be learned from the observed data $Y$, for example, through NMF. 
We use iterated expectation to expand the identified statistical estimand that relies only on latent outcomes $L$ (Equation \ref{eq:ident}) to an estimand that can be estimated from the observed data (Equation \ref{eq:est}).
\begin{align}
  \mathbb E[ L(t)] &= \mathbb E[ L|T = t] & \text{Identified} \label{eq:ident}\\
  &= \mathbb E_{Y}[\mathbb E[ L|T = t, Y]]. & \text{Estimable} \label{eq:est}
\end{align}

We define a latent outcome model $\ell_{A, t}$ as a function trained using an algorithm $A$ such that the \textit{learned latent outcome} $\ell_{A, t}(Y_i, T_i)$ is an estimate of the inner expectation, $\mathbb E[ L|T = t, Y = Y_i]$ (Equation \eqref{eq:est}), depending on observed data $Y_i$ and observed treatment assignment $T_i$. This is analogous to an outcome model, often denoted $\mu_t(X_i)$, as used in g-computation or augmented inverse propensity weighting (AIPW) estimation \citep{hernan2010causal}. In $\ell_{A, t}(Y_i, T_i)$, the subscript $t$ refers to the treatment assignment of the estimand, while the input $T_i$ is the observed treatment level for subject $i$. This notation underscores that the estimation procedure may be different depending on whether $T_i = t$, that is, whether $Y_i(t)$ is observed. For some algorithms, only $\ell_{A, T_i}(Y_i, T_i)$ under the observed treatment level $T_i$ may be learned, while others may learn $\ell_{A, 1-T_i}(Y_i, T_i)$ as well.

If the latent outcome model $\ell_{A, t}(Y_i, T_i)$ provides unbiased estimates of the inner expectation $\mathbb E[L|T = t, Y = Y_i]$ from Equation \eqref{eq:est}, they can be averaged across samples within treatment groups to yield unbiased estimates of the outer expectation $\mathbb E[L|T = t]$ from Equation \eqref{eq:ident} and further for an unbiased estimate of the ATE, $\mathbb E[L|T = 1] - \mathbb E[L|T = 0]$.

However, even under the assumption of no interference in the data generating process of $L$, \textit{the latent outcome model may still depend on the full treatment program $\mathbf T$} through its training, resulting in biased estimates $\ell_{A, t}(Y_i, T_i)$. In this setting, no interference can be thought of as a desired property that applies to the latent outcome model. We define this as the property of \textit{no learning-induced interference}: the learned latent outcome $\ell_{A, t}(Y_i, T_i)$ of individual $i$ should not depend on the treatments of others, $\mathbf T_{-i}$ (Property \ref{property:no_means_interf}).

\begin{property}[No learning-induced interference]
\label{property:no_means_interf}
Let $\ell_{A, t}$ be a latent outcome model using algorithm $A$ and trained on observed data $\{\mathbf T, \mathbf Y(\mathbf T)\}$, and let $\ell^*_{A, t}$ be a latent outcome model from the same algorithm and trained on data $\{\mathbf T^*, \mathbf Y(\mathbf T^*)\}$ generated by a counterfactual treatment program $\mathbf T^*$. We say that algorithm $A$ satisfies \textit{no learning-induced interference} if, for any unit $i$ and any pair $\mathbf T, \mathbf T^*$ such that $T^*_i = T_i$, we have $\ell_{A, t}(Y_i, T_i) = \ell^{*}_{A, t}(Y_i, T_i)$.
\end{property}

If the latent outcome model is trained on all observations, as it is in the All Data algorithm (as in \citet{knox2022testing}), this property is typically not met as the learned latent outcomes inherently depend on all treatments. Figure \ref{figure:dag1} visualizes this concept with a representative pair of subjects $i$ and $j$. Learning the factor model induces a clear path from the treatment of one subject, $T_i$, to the learned latent outcome of another, $\ell_{A, T_j}(Y_j, T_j)$. The primary goal of our work is to reduce learning-induced interference by minimizing the magnitude of the effect $\mathbf Y \to \hat{\boldsymbol \lambda}$, as discussed in detail in Section \ref{sec:methods}.

To avoid learning-induced interference, \cite{egami2022make} recommends developing the factor model on a subset of the data and estimating causal effects with the held-out data. We refer to this approach as the \textit{Random Split algorithm} and include it in our simulation and data application comparisons. However, we argue that this approach only avoids interference within the held-out data, but still allows the treatments of the factor model subset to affect the learned latent outcomes in the held-out data. We will show in Section \ref{sec:simulations} that indirect effects quantifying learning-induced interference are not improved by this approach. Further, splitting the data in this way increases variability in the factor model and downstream in the causal estimates.

The idea of learning-induced interference can be extended beyond latent outcomes more generally to scenarios with measurement error. Even if a true data generating process is free of interference, practitioners must make sure that one subject's treatment does not affect the error on another subject's measured outcomes. For example, if an outcome is the weight of produce or livestock, and treated subjects weigh substantially more than untreated subjects, a scale may become miscalibrated from the treated heavy subjects, therefore impacting the measurement error on later subjects. 
Learning-induced interference can, in fact, be present in standard outcome models used for g-computation or AIPW estimation. However, we expect to see a much larger magnitude of learning-induced interference in our setting because of the interdependent structure and complexity of factor models.

\usetikzlibrary{arrows.meta, positioning, calc}
\usetikzlibrary{shapes.geometric}

\begin{figure}[t]
    \centering
    \begin{tikzpicture}[
        roundnode/.style={circle, draw=black, thick, minimum size=1.4cm, font=\large, fill=white},
        roundnodesmall/.style={circle, draw=black, thick, minimum size=0.1cm, font=\large, fill=white},
        estimated/.style={ellipse, draw=black, fill=gray!20, thick, minimum height=1.4cm, font=\small, inner sep=6pt, align=center},
        estimated_circle/.style={circle, draw=black, fill=gray!20, thick, minimum height=1.4cm, font=\large, inner sep=6pt, align=center},
        estimatedsmall/.style={circle, draw=black, fill=gray!20, thick, minimum size=0.1cm, font=\large},
        arrow/.style={-Stealth, thick},
        redarrow/.style={-Stealth, thick, red},
        dashedarrow/.style={-Stealth, thick, dashed, gray},
      ]
    
      \node[roundnode] (Ti) at (0,0) {$T_i$};
      \node[estimated] (Li) at (3,0) {$\ell_{A, T_i}(Y_i, T_i)$};
      \node[roundnode] (Yi) at (6,0) {$Y_i$};
    
      \node[roundnode] (Tj) at (0,-2) {$T_j$};
      \node[estimated] (Lj) at (3,-2) {$\ell_{A, T_j}(Y_j, T_j)$};
      \node[roundnode] (Yj) at (6,-2) {$Y_j$};
    
      \node[estimated_circle] (lambda) at (3,2.4) {$\hat {\boldsymbol \lambda}$};
      \node[roundnode] (Y) at (7.7,-1) {$\mathbf{Y}$};
    
      \draw[redarrow] (Ti) -- (Li);
      \draw[arrow] (Li) -- (Yi);
      \draw[redarrow] (Tj) -- (Lj);
      \draw[arrow] (Lj) -- (Yj);
    
      \draw[arrow] (lambda) -- (Yi);
      \draw[arrow] (lambda) -- (Yj);
      \draw[dashedarrow] (Y) to[out=90,in=0] (lambda);
    
      \draw[dashedarrow] (Yi) to[out=200,in=-20] (Li);
      \draw[dashedarrow] (Yj) to[out=200,in=-20] (Lj);
    
      \node at ($(Yi)!0.4!(Y)$) [rotate=-30, anchor=south west, font=\Large] {$\in$};
      \node at ($(Yj)!0.4!(Y)$) [rotate=30, anchor=north west, font=\Large] {$\in$};
    
      \begin{scope}[xshift=12cm, yshift=-0.5cm]
      \node[rectangle, draw=black, thick, fill=white, inner sep=8pt] (leg) at (0,0) {
        \begin{tabular}{@{}m{1cm} m{3.5cm}@{}}
          \tikz{\draw[arrow] (0,0) -- (0.6,0);} & Causal effect \\
          \tikz{\draw[redarrow] (0,0) -- (0.6,0);} & Causal effect of interest \\
          \tikz{\draw[dashedarrow] (0,0) -- (0.6,0);} & Estimation \\
          \tikz{\node[estimatedsmall] at (0,0) {}; } & Estimated \\
          \tikz{\node[roundnodesmall] at (0,0) {}; } & Observable \\
        \end{tabular}
      };
    \end{scope}
    
    \end{tikzpicture}
    \caption{\textbf{DAG with estimates in place of unobservable variables}, where $\ell_{A, T_i}(Y_i, T_i)$ is the learned latent outcome, or an estimate of $\mathbb E[ L|T = t, Y = Y_i]$, under algorithm $A$. This DAG demonstrates learning-induced interference via a path from $T_i$ to $\ell_{A, T_j}(Y_j, T_j)$ and from $T_j$ to $\ell_{A, T_i}(Y_i, T_i)$ for each $i \ne j$. Minimizing the magnitude of the effect $\mathbf Y \to \hat{\boldsymbol \lambda}$ to reduce learning-induced interference is the target of our work.}
    \label{figure:dag1}
\end{figure}
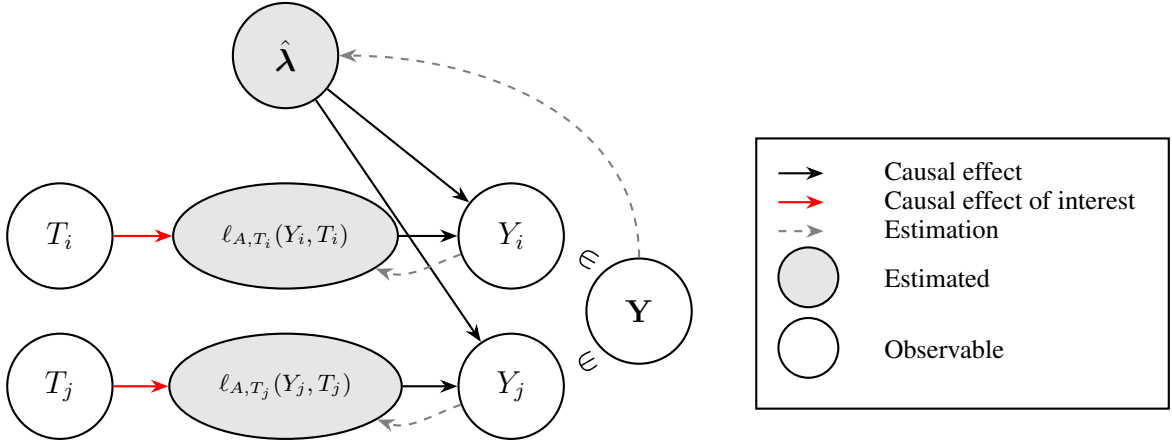

\subsection{Quantifying learning-induced interference with indirect effects}\label{sec:indirect_effects}

There is an existing body of work on causal inference under interference. There are generally two points of view: interference is something to be avoided with study design (e.g., plot arrangement in agricultural studies) \citep{neyman1923application, rosenbaum_interference_2007}, or interference is of scientific interest and something to be estimated (e.g., herd immunity in vaccine trials) \citep{hudgens_toward_2008, halloran_dependent_2016}. This work falls more closely within the first viewpoint---we aim to develop an algorithm to avoid learning-induced interference. 
However, we borrow causal estimands from the second viewpoint to use as metrics when comparing algorithms in simulation studies. 

Adapting the general notation of \cite{hudgens_toward_2008}, indirect effects measure the change in an untreated individual's outcome as the treatment assignment mechanism used to treat others changes. For each effect, we first introduce it as it appears in existing literature---as a causal estimand with respect to the true latent outcome $L$. To quantify learning-induced interference, we re-define it as a statistical metric in terms of the learned latent outcome $\ell_{A, t}(Y_i, T_i)$. Consider $\pi$ as a treatment assignment mechanism, which at its simplest is the proportion treated in a completely randomized design. 

The definition of indirect effects requires the concept of individual average potential outcomes (IAPOs). The IAPO $\bar{L}_i(t|\pi)$ is the expectation of individual $i$'s true latent outcome when they are given the specified treatment $T_i = t$, averaged over all possible treatment assignments of other individuals $\mathbf T_{-i}$ under treatment assignment mechanism $\pi$ (Equation \ref{eq:IAPO}). Under the causal assumption of no interference in the data generating process, $\bar{L}_i(t|\pi)$ does not depend on $\pi$. To quantify learning-induced interference, we adapt this causal estimand into a statistical metric, the individual average learned latent outcome (IALLO), which takes the same form but as an average of learned latent outcomes $\ell_{A, t}(Y_i, T_i)$ conditional on $\ell_{A, t}$ having been trained on a given treatment program (Equation \ref{eq:IAMO}).

\noindent
\begin{minipage}[t]{0.48\textwidth}
\begin{equation}
\bar{L}_i(t|\pi) = \mathbb E_{\mathbf{T}_{-i} \sim \pi}\left[ L_i(T_i = t, \mathbf{T}_{-i})\right] \label{eq:IAPO}
\end{equation}
\end{minipage}
\hfill
\begin{minipage}[t]{0.5\textwidth}
\begin{equation}
\bar{\ell}_{A, i}(t|\pi) = \mathbb E_{\mathbf{T}_{-i} \sim \pi}\left[ \ell_{A, t}(Y_i, T_i) | T_i = t, \mathbf T_{-i}\right] \label{eq:IAMO}
\end{equation}
\end{minipage}

The individual average indirect effect, $\widebar{IE}_{L,i}(\pi, \pi')$, is the difference in untreated IAPOs between two treatment assignment mechanisms $\pi$ and $\pi'$ (Equation \ref{eq:IAIE}). Averaging this quantity across all samples yields the population average indirect effect, $\widebar{IE}_{L}(\pi, \pi')$ (Equation \ref{eq:PAIE}). Similarly, under a given algorithm $A$, the learning-induced individual average indirect effect, $\widebar{IE}_{\ell, A, i}(\pi, \pi')$, is the difference in untreated IALLOs (liIAIE, Equation \ref{eq:mIAIE}), and its average over all samples is the learning-induced population average indirect effect, $\widebar{IE}_{\ell, A}(\pi, \pi')$, (liPAIE, Equation \ref{eq:mPAIE}).

\noindent
\begin{minipage}[t]{0.48\textwidth}
\begin{equation}
\widebar{IE}_{L, i}(\pi, \pi') = \bar{L}_i(0|\pi) - \bar{L}_i(0|\pi') 
\label{eq:IAIE}
\end{equation}
\end{minipage}
\hfill
\begin{minipage}[t]{0.48\textwidth}
\begin{equation}
\widebar{IE}_{\ell, A, i}(\pi, \pi') = \bar{\ell}_{A, i}(0|\pi) - \bar{\ell}_{A, i}(0|\pi') 
\label{eq:mIAIE}
\end{equation}
\end{minipage}

\noindent
\begin{minipage}[t]{0.48\textwidth}
\begin{equation}
\widebar{IE}_{L}(\pi, \pi') = \frac{1}{N}\sum_{i = 1}^N \widebar{IE}_{L, i}(\pi, \pi') 
\label{eq:PAIE}
\end{equation}
\end{minipage}
\hfill
\begin{minipage}[t]{0.48\textwidth}
\begin{equation}
\widebar{IE}_{\ell, A}(\pi, \pi') = \frac{1}{N}\sum_{i = 1}^N \widebar{IE}_{\ell, A, i}(\pi, \pi') 
\label{eq:mPAIE}
\end{equation}
\end{minipage}

If there is no interference (regular or learning-induced), the corresponding indirect effects are always zero. However, the converse does not hold: indirect effects at zero do not necessarily imply the absence of interference. Still, when indirect effects are zero, any remaining interference does not affect the ATE, making it a valid and meaningful measure of the causal effect of interest. While the causal indirect effects are unmeasurable (and known to be zero as we assume no interference on the data generating process), we are able to compute the learning-induced statistical metrics in simulation studies to quantify the degree of interference across various algorithms.





\section{Motivating example in cancer mutational signatures analysis}

Although this work has general applications to factor models and other applications of Poisson NMF, we focus on cancer mutational signatures for our motivating example, simulation studies, and data application. Mutational signatures analysis models a tumor's mutational landscape as a composition of multiple mutational processes acting simultaneously \citep{alexandrov2013signatures}. In this context, $\mathbf Y_i$ is a vector of mutation counts for the individual, or tumor genome, $i$ across $D$ mutation types. NMF is used with a Poisson likelihood to estimate a signatures matrix $\boldsymbol \lambda$, where each column $\lambda_k$ is a probability distribution over mutation types that sums to $1$, and a contributions matrix $\mathbf L$, where each column $L_i$ indicates how many mutations in genome $i$ are attributed to each of the $K$ mutational signatures. A popular set of reference signatures is provided by the Catalog of Somatic Mutations in Cancer (COSMIC) database \citep{tate2019cosmic}. While these signatures are useful for comparison, they were estimated from data and therefore cannot be treated as comprehensive or ground truth.

\begin{figure}
  \centering
  \includegraphics[width=0.8\linewidth]{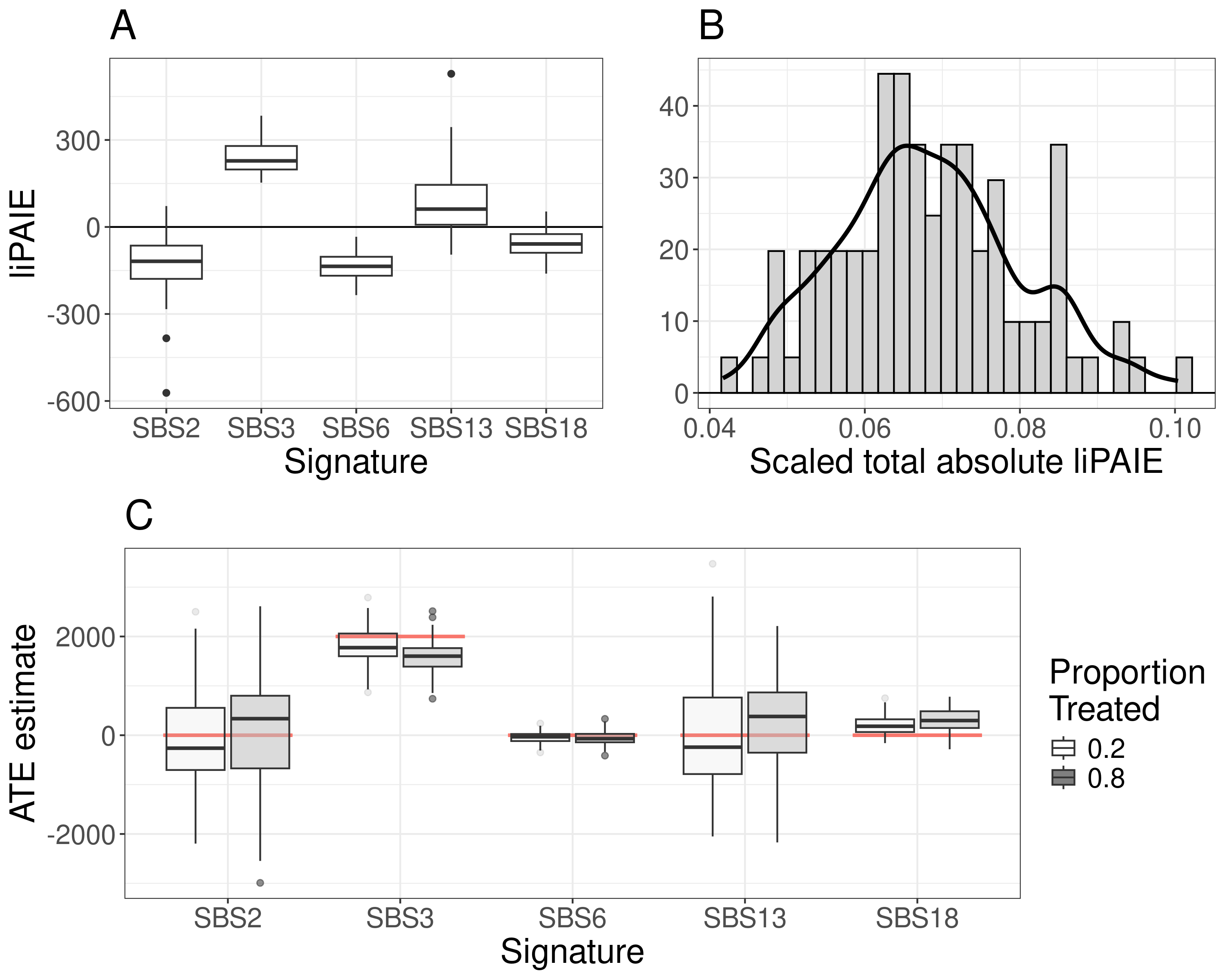}
  \caption{\textbf{Motivational example: indirect effects of the All Data algorithm} across 100 simulated datasets of 100 individuals each. \textbf{A) Learning-induced population average indirect effects (liPAIEs)} for 5 cancer mutational signatures. This represents the expected change in a single dimension of an untreated individual's learned latent outcome, or number of mutations attributed to the given signature, if other subjects change from 20\% treated to 80\% treated. Without learning-induced interference, these values will be centered at 0. \textbf{B) Sum of absolute liPAIEs per sample}, rescaled by twice the number of mutations per sample (any change is counted by liPAIE of both old and new signature attribution). This represents the proportion of an individual's mutations whose attribution changes due to the shift of other subjects from 20\% treated to 80\% treated. \textbf{C) Bootstrapped mean ATE estimates} with either 20\% treated (filled in white) or 80\% treated (filled in grey). Under no learning-induced interference, we expect the same ATE estimates regardless of the proportion treated.}
  \label{fig:motivation}
\end{figure}

In this section, we provide an example illustrating the repercussions of ignoring learning-induced interference. The data for this example are simulated in the context of cancer mutational signatures according to Section \ref{sec:simulations}, where the latent outcome holds the number of mutations attributed to each of the five mutational signatures: SBS2, SBS3, SBS6, SBS13, and SBS18. Unlike abstract latent dimensions, mutational signatures have interpretable units, allowing the liPAIE to be understood directly as the number (or if normalized the proportion) of mutations whose attribution changes due to a change in treatment assignment mechanism. The results in this section show that learning-induced interference is not just a theoretical problem, but a quantifiable issue that can have large consequences on learned latent outcomes and estimated causal effects in practice.

We compute learning-induced population average indirect effects $\widebar{IE}_{\ell}(\pi = 0.2, \pi' = 0.8)$ comparing scenarios in which 20\% versus 80\% of individuals are treated. The baseline All Data approach estimates factors on all observations and estimates causal effects as a difference of treatment-group means on the learned latent outcomes, again using all observations. The magnitude of the change in the learned latent outcome due to this shift in treatment assignment strategy is on the order of hundreds of mutations (Figure \ref{fig:motivation}A). The proportion of mutations whose attribution changes is centered around 6.5\% and exceeds 10\% in some cases (Figure \ref{fig:motivation}B).

The downstream, and perhaps more meaningful, effect of learning-induced interference is on the average treatment effect (ATE) estimates. As the percent of treated individuals increases from 20\% to 80\%, the estimated ATEs of SBS2, SBS13, and SBS18 increase, while the estimated ATEs of SBS3 and SBS6 decrease (Figure \ref{fig:motivation}C). The change in ATE estimates for signatures SBS2 and SBS13 ranges to over 500 mutations. These estimates are biased in both of the counterfactual worlds, often in different directions.

\section{Methods}
\label{sec:methods}

\usetikzlibrary{arrows.meta, positioning, calc}

\begin{figure}[ht!]
\centering
\begin{tikzpicture}[
    roundnode/.style={circle, draw=black, thick, minimum size=1.4cm, font=\large, fill=white},
    roundnodesmall/.style={circle, draw=black, thick, minimum size=0.1cm, font=\large, fill=white, scale = 0.7},
    estimated/.style={ellipse, draw=black, fill=gray!20, thick, minimum height=1.4cm, font=\small, inner sep=6pt, align=center},
    estimated_circle/.style={circle, draw=black, fill=gray!20, thick, minimum height=1.4cm, font=\large, inner sep=6pt, align=center},
    estimatedsmall/.style={circle, draw=black, fill=gray!20, thick, minimum size=0.1cm, font=\large, scale = 0.7},
    boxnode/.style={rectangle, draw=black, fill=gray!20, thick, minimum height=1.2cm, minimum width=1.8cm, font=\large, align=center},
    boxnodesmall/.style={rectangle, draw=black, fill=gray!20, thick, minimum height=0.1cm, minimum width=0.3cm, font=\large, align=center, scale = 0.95},
    arrow/.style={-Stealth, thick},
    redarrow/.style={-Stealth, thick, red},
    dashedarrow/.style={-Stealth, thick, dashed, gray},
  ]

  \node[roundnode] (Ti) at (-1,0) {$T_i$};
  \node[estimated] (Li) at (2,0) {$\ell_{\text{IS}, T_i}(Y_i, T_i)$}; 
  \node[roundnode] (Yi) at (5,0) {$Y_i$};

  \node[roundnode] (Tj) at (-1,-2.2) {$T_j$};
  \node[estimated] (Lj) at (2,-2.2) {$\ell_{\text{IS}, T_j}(Y_j, T_j)$};
  \node[roundnode] (Yj) at (5,-2.2) {$Y_j$};

  \node[estimated_circle] (lambda) at (2,3.3) {$\hat{\boldsymbol \lambda}$};
  \node[estimated_circle] (Y0) at (5,3.8) {$\tilde{\mathbf{Y}}_0$};
  \node[estimated_circle] (Y1) at (5,2.2) {$\tilde{\mathbf{Y}}_1$};
  \node[boxnode] (fimp) at (8,3.3) {$f_{\text{IMP}}$};
  \node[roundnode] (Y) at (6.7,-1.1) {$\mathbf{Y}$};

  \draw[redarrow] (Ti) -- (Li);
  \draw[arrow] (Li) -- (Yi);
  \draw[redarrow] (Tj) -- (Lj);
  \draw[arrow] (Lj) -- (Yj);
  \draw[arrow] (lambda) -- (Yi);
  \draw[arrow] (lambda) -- (Yj);

  \draw[dashedarrow] (Y) -- (fimp);
  \draw[dashedarrow] (fimp) -- (Y0);
  \draw[dashedarrow] (fimp) -- (Y1);
  \draw[dashedarrow] (Y0) -- (lambda);
  \draw[dashedarrow] (Yi) to[out=200,in=-20] (Li);
  \draw[dashedarrow] (Yj) to[out=200,in=-20] (Lj);

  \node at ($(Yi)!0.4!(Y)$) [rotate=-25, anchor=south west, font=\Large] {$\in$};
  \node at ($(Yj)!0.4!(Y)$) [rotate=25, anchor=north west, font=\Large] {$\in$};

  \begin{scope}[xshift=12cm, yshift=-0.5cm]
  \node[rectangle, draw=black, thick, fill=white, inner sep=8pt] (leg) at (0,0) {
    \begin{tabular}{@{}m{1cm} m{3.5cm}@{}}
      \tikz{\draw[arrow] (0,0) -- (0.6,0);} & Causal effect \\
      \tikz{\draw[redarrow] (0,0) -- (0.6,0);} & Causal effect of interest \\
      \tikz{\draw[dashedarrow] (0,0) -- (0.6,0);} & Estimation \\
      \tikz{\node[estimatedsmall] at (0,0) {}; } & Estimated \\
      \tikz{\node[boxnodesmall] at (0,0) {};} & Estimated Function\\
      \tikz{\node[roundnodesmall] at (0,0) {}; } & Observable \\
    \end{tabular}
  };
  \end{scope}

\end{tikzpicture}
\caption{\textbf{DAG with Impute and Stabilize estimates in place of unobservable variables}, where $\ell_{\text{IS}, T_i}(Y_i, T_i)$ is the learned latent outcome, or an estimate of $\mathbb E[ L|T = t, Y = Y_i]$, under the Impute and Stabilize (IS) algorithm. Direct learning-induced interference via $\mathbf Y \to \hat{\boldsymbol \lambda}$ seen in Figure \ref{figure:dag1} has been replaced with a reduced, more indirect form of learning-induced interference through imputation $f_{\text{IMP}}$ and a stabilized matrix factorization on $\tilde{\mathbf{Y}}(0)$ alone.}
\label{figure:dag2}
\end{figure}
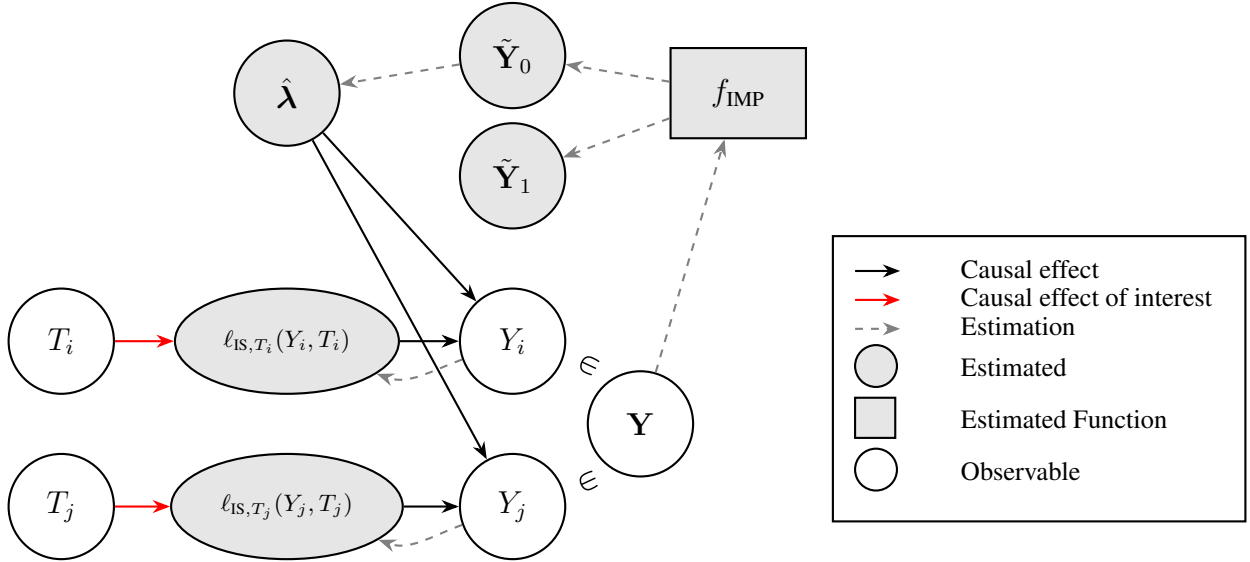

\subsection{Novel algorithm: Impute and Stabilize}

Our primary goal with this novel algorithm is to reduce the effect of learning-induced interference. This is accomplished through stabilization of the factor model by providing an input that is less dependent on treatment---specifically, untreated samples alone. 
Intuitively, the All Data algorithm is subject to high levels of learning-induced interference because changing a single individual's treatment changes an entire column of $\mathbf Y$, that is, the factor model input is highly sensitive to changes in treatment. This is a change of large magnitude with large expected effects on the estimated $\hat{\boldsymbol \lambda}$. Within the subset of untreated individuals $\mathbf Y_{\{i:T_i = 0\}}$, we do not have as severe treatment-driven variability because changing a single individual's treatment simply adds or removes an individual from the factor model input, and we expect smaller effects on estimated $\hat{\boldsymbol \lambda}$. An NNLM can be used on the remaining $\mathbf Y_{\{i:T_i = 1\}}$ with fixed $\hat{\boldsymbol \lambda}$ to learn the latent outcomes under treatment. This describes the mechanism of the \textit{stabilization} strategy.

However, fitting the factor model on a subset of data increases variability due to a smaller sample size. We address this by combining the stabilization strategy with \textit{imputation}. Although $Y$ is not the outcome of interest, it is a post-treatment variable, and we can estimate its unobserved potential outcomes, $\mathbf Y(1-\mathbf T)$, with imputations $\tilde{\mathbf Y}_{1-\mathbf T}$. Observed and imputed data can be combined to construct a matrix $\tilde{\mathbf Y}_{\mathbf 0}$ of the original sample size such that each column $\tilde{\mathbf Y}_{\mathbf 0,i}$ is set to the observed $Y_i$ if $T_i = 0$ and set to the imputed $\tilde Y_{1-\mathbf T, i}$ if $T_i = 1$, and similarly for $\tilde{\mathbf Y}_{\mathbf 1}$. The factor model can be fit on this $\tilde{\mathbf Y}_{\mathbf 0}$ to learn $\hat {\boldsymbol \lambda}$ and untreated latent outcomes, and an NNLM can be used on the remaining $\tilde{\mathbf Y}_{1}$ with fixed $\hat{\boldsymbol \lambda}$ to learn treated latent outcomes. 

By integrating the \textit{Impute} and \textit{Stabilize} steps, this algorithm improves estimation of the causal effect in three ways. First, as compared to stabilization alone, the increased sample size reduces variability in the factor model. Second, the full $\tilde{\mathbf Y}_{\mathbf 0}$ matrix is less sensitive to changes in treatment than either the observed data $\mathbf Y$ or the subset $\mathbf Y_{\{i:T_i = 0\}}$ as input to the factor model. Here, changing a single individual's observed treatment typically induces small perturbations in the estimated imputations, with a smaller expected effect on $\hat{\boldsymbol \lambda}$. Third, the imputation strategy allows for paired, or within-sample, contrasts $\ell_{\text{IS}, 1}(Y_i, T_i)$ versus $\ell_{\text{IS}, 0}(Y_i, T_i)$, improving efficiency in estimated ATEs, our secondary goal in this work. Figure \ref{figure:dag2} visualizes how imputation ($\mathbf Y \to f_{\text{IMP}} \to \tilde {\mathbf Y}_0, \tilde {\mathbf Y}_1$) and stabilization ($\tilde{\mathbf Y}_0 \to \hat{\boldsymbol \lambda}$) work together to reduce the magnitude of interference, or the strength of the path $\mathbf Y \to \hat{\boldsymbol \lambda}$. While it is clear that this path has not been removed, our simulation studies confirm that it has been substantially reduced in magnitude.

\subsection{Imputation function $f_{\text{IMP}}$ for Poisson data}\label{sec:fimp}

The success of the Impute and Stabilize algorithm requires accurate imputation, which depends on distributional assumptions of $Y$. Recall that the observed data are assumed to follow a Poisson distribution: ${Y}_i(t) \sim \text{Poisson}(\boldsymbol{\lambda} {L}_i(t))$, where $\mathbb{E}[{Y}_i(t)] = \text{Var}[{Y}_i(t)]$. This implies that individuals with higher baseline rates exhibit greater variability in their observed data, regardless of treatment. Consequently, assuming a constant treatment effect on $Y$ is inappropriate. Instead, we assume that treatment effects on $Y$ are constant on the square-root transformed scale, which stabilizes the variance of Poisson-distributed data \citep{bartlett1936square, anscombe1948transformation}. The imputation proceeds in three steps:

\begin{enumerate}
  \item \textbf{Variance stabilization.} Apply the square-root transformation:
  \[
  \mathbf{Y}^{\text{vst}} = \sqrt{\mathbf{Y}}.
  \]

  \item \textbf{Imputation via difference-in-means.} Estimate the treatment effect on the stabilized scale:
  \begin{align*}
  \boldsymbol\psi_{\mathbf{Y}}^{\text{vst}} &= \mathbb E\left[Y^{\text{vst}}(1) - Y^{\text{vst}}(0)\right],\\
  \hat{\boldsymbol\psi}_{\mathbf{Y}}^{\text{vst}} &= \frac{1}{N_1} \sum_{i: T_i = 1} Y^{\text{vst}}_i 
  - \frac{1}{N_0} \sum_{i: T_i = 0} Y^{\text{vst}}_i.
  \end{align*}
  Impute unobserved potential outcomes, using observed $Y^{\text{vst}}_i$ as an estimate of $\mathbb E\left[Y^{\text{vst}}_i(T_i)\right]$:
  \begin{align*}
    \mathbb E\left[Y^{\text{vst}}_i(1-T_i)\right] &= \mathbb E\left[Y^{\text{vst}}_i(T_i)\right] + (1 - T_i) \cdot {\boldsymbol\psi}_{\mathbf{Y}}^{\text{vst}} - T_i \cdot {\boldsymbol\psi}_{\mathbf{Y}}^{\text{vst}},\\
    \tilde{Y}^{\text{vst}}_{1-T_i, i} &= Y^{\text{vst}}_i + (1 - T_i) \cdot \hat{\boldsymbol\psi}_{\mathbf{Y}}^{\text{vst}} - T_i \cdot \hat{\boldsymbol\psi}_{\mathbf{Y}}^{\text{vst}}.
  \end{align*}

  \item \textbf{Back-transformation.} The back-transformation expression is based on the equality 
  
  \begin{equation*}
    \mathbb{E}\left[Y_i(1-T)\right] = \mathbb{E}\left[\left(Y_i^{\text{vst}}(1-T)\right)^2\right] = \mathbb{E}\left[Y^{\text{vst}}_i(1-T)\right]^2 + \text{Var}\left[Y^{\text{vst}}_i(1-T)\right].
  \end{equation*}

  For any Poisson-distributed variable $Y_i$, the approximation that $\text{Var}[\sqrt{Y_i}] \approx 1/4$ is valid for reasonably sized $\mathbb E[Y_i] \gtrsim 5$ \citep{bartlett1936square, anscombe1948transformation}, giving an approximation for $\text{Var}[Y^{\text{vst}}_i] \approx 1/4$. Based on the definition of $\hat{\boldsymbol\psi}_{\mathbf{Y}}^{\text{vst}}$, we get 
  \begin{equation*}
    \text{Var}[\hat{\boldsymbol\psi}_{\mathbf{Y}}^{\text{vst}}] \approx \frac{1}{4} \left( \frac{1}{N_1} + \frac{1}{N_0} \right).
  \end{equation*}
  
  Finally, we assume $\text{Cov}(Y_i^{\text{vst}}, \hat{\boldsymbol\psi}_{\mathbf{Y}}^{\text{vst}}) \approx 0$, which is reasonable for large $N$. Therefore, we can conclude that
  \begin{equation*}
    \text{Var}[Y^{\text{vst}}_i(1-T)] = \text{Var}\left[Y^{\text{vst}}_i \pm \hat{\boldsymbol\psi}_{\mathbf{Y}}^{\text{vst}} \right] \approx \frac{1}{4} \left( 1 + \frac{1}{N_1} + \frac{1}{N_0} \right).
  \end{equation*}

  This yields the final back-transformation expression 
  \begin{equation*}
    \tilde{Y}_{1-T_i, i} = \left( \tilde{Y}^{\text{vst}}_{1-T_i, i} \right)^2 + \frac{1}{4} \left( 1 + \frac{1}{N_1} + \frac{1}{N_0} \right).
  \end{equation*}
\end{enumerate}

We chose to use a theoretically derived estimate of $\text{Var}\left[Y^{\text{vst}}_i(1-T)\right]$ instead of a value estimated from the data to avoid further learning-induced interference.

\subsection{Baseline and ablation algorithms}

\renewcommand{\arraystretch}{3}
\begin{table}[]
  \centering
  \begin{tabular}{|c|c|c|c|c|}
    \hline
    \textbf{Algorithm} & \makecell{\textbf{Preprocessing}\\\textbf{and definitions}} & \makecell{\textbf{NMF input} \\ $\to$ learns what $\ell$} & \makecell{\textbf{NNLM input} \\ $\to$ learns what $\ell$} & \textbf{ATE estimator $\hat{\boldsymbol \psi}_L$}\\

    \hline
    Oracle & $\ell_{\text{O}, t}(Y_i, T_i) := L_i(t)$ & - & - & \makecell{mITE} \\
    \makecell{Observed\\Outcome} & \makecell{$\ell_{\text{OO}, t}(Y_i, T_i) := $ \\ $\frac{1}{N_t}\sum_{i:T_i = t}L_i(T_i)$} & - &- & \makecell{DM}\\
    \hline
    All Data & - & \makecell{$\mathbf Y$ \\ $\to \ell_{\text{AD}, T_i}(Y_i, T_i)$} & - & DM \\
    \makecell{Random\\Split} & \makecell{Sample indices $S$ \\ $||S|| = \lceil N/2\rceil$} & $\mathbf Y_S$ & \makecell{$\mathbf Y_{/S}$ \\ $\to \ell_{\text{RS}, T_i}(Y_i, T_i), i\notin S$} & DM among $i \notin S$\\
    \hline
    Impute & \makecell{Impute $\tilde{\mathbf Y}_{1 - \mathbf{T}}$ \\ with $f_{\text{IMP}}$} & \makecell{$\mathbf Y$ \\ $\to \ell_{\text{I}, T_i}(Y_i, T_i)$} & \makecell{$\tilde{\mathbf Y}_{1 - \mathbf T}$ \\ $\to \ell_{\text{I}, 1-T_i}(Y_i, T_i)$} & mITE \\
    Stabilize & - & \makecell{$\mathbf Y_{\{i:T_i = 0\}}$ \\ $\to \ell_{\text{S}, T_i}(Y_i, T_i), i:T_i = 0$} & \makecell{$\mathbf Y_{\{i:T_i = 1\}}$ \\ $\to \ell_{\text{S}, T_i}(Y_i, T_i), i:T_i = 1$} & DM\\
    \makecell{Impute and\\Stabilize} & \makecell{Impute $\tilde{\mathbf Y}_{1 - \mathbf{T}}$ \\ with $f_{\text{IMP}}$} & \makecell{$\tilde{\mathbf Y}_0$ \\ $\to \ell_{\text{IS}, 0}(Y_i, T_i)$} & \makecell{$\tilde{\mathbf Y}_1$ \\ $\to \ell_{\text{IS}, 1}(Y_i, T_i)$} & mITE\\
    \hline
  \end{tabular}
  \caption{\textbf{Algorithm definitions: preprocessing steps, how latent outcome models $\ell_{A, t}$ are learned as using estimated factor weights from NMF and possibly NNLM, and how estimates are combined into an ATE estimator $\hat {
  \boldsymbol\psi}_L$.} $\mathbf Y$ is the full observed data matrix and $\mathbf Y_I$ for any set of indices $I$ is a subset of the data matrix corresponding to columns with indices $i \in I$. A tilde $\tilde{\mathbf Y}$ indicates at least some values of the matrix have been replaced with imputations, where $\tilde{\mathbf Y}_{1-\mathbf T}$ is entirely imputed and $\tilde{\mathbf Y}_t$ is a combination of observed (for $i$ where $T_i = t$) and imputed (for $i$ where $T_i \ne t$) values. Recall that $\ell_{A, t}(Y_i, T_i)$ refers to an estimate of $\mathbb E[L|T_i = t, Y = Y_i]$ using algorithm $A$. In some cases (Oracle, Impute, Impute and Stabilize), we are able to compute $\ell_{A, 1-T_i}(Y_i, T_i)$ and can utilize paired contrasts in a mean individual treatment effect (mITE) estimator of the form $\frac{1}{N}\sum_{i = 1}^N\left(\ell_{A, 1}(Y_i, T_i) - \ell_{A, 0}(Y_i, T_i)\right)$. In all other algorithms, only $\ell_{A, T_i}(Y_i, T_i)$ for observed treatment level $T_i$ can be computed, so we must use a difference of means (DM) estimator of the form $\frac{1}{N_1}\sum_{i: T_i = 1}\ell_{A, T_i}(Y_i, T_i) - \frac{1}{N_0}\sum_{i: T_i = 0}\ell_{A, T_i}(Y_i, T_i)$. The first set of algorithms are only possible to use in simulation studies. The second set of algorithms are currently used or suggested in literature. The final set of algorithms are developed in this paper: two ablations and our complete novel Impute and Stabilize algorithm. [NMF: nonnegative matrix factorization. NNLM: nonnegative linear model, equal to NMF with a fixed factor matrix. DM: difference of means. mITE: mean individual treatment effect. $f_{\text{IMP}}$: imputation function (see Section \ref{sec:fimp}).]}
  \label{tab:algorithms}
\end{table}

We compare our approach with two baselines. The first is the All Data approach, which uses all observations $\mathbf Y$ to perform matrix decomposition to estimate factors $\hat{\boldsymbol \lambda}$ and learned latent outcomes $\ell_{\text{AD}, T_i}(Y_i, T_i)$, then again uses all observations to estimate causal effects with difference of means on $\ell_{\text{AD}, T_i}(Y_i, T_i)$. 

Second, we consider the Random Split approach suggested by \citet{egami2022make}, which identifies a random subset of 50\% of indices $S$, uses $\mathbf Y_S$ as input to matrix decomposition to estimate $\hat{\boldsymbol \lambda}$, uses an  NNLM on the remaining data $\mathbf Y_{/S}$ to learn latent outcomes $\ell_{\text{RS}, T_i}(Y_i, T_i)$ for $i \notin S$, and finally estimates the ATE with difference of means on these estimates.

In simulation settings, we define an Oracle approach, which assumes $L$ is not latent and that we can observe both potential outcomes. Using true latent potential outcomes $\ell_{\text{O}, t}(Y_i, T_i) = L_i(t)$, we compute the mean of individual treatment effects (ITEs). The Oracle is never possible to attain as both potential outcomes can never be observed, even if ${L}$ was not a latent variable. 

Also in simulation settings, we define the Observed Outcome algorithm, which assumes $L$ is not latent, but we can only observe $L_i(T_i)$ for observed treatment $T_i$. We compute the difference in means estimator on the true latent outcomes. Here, group-specific means are used as the latent outcome model such that $\ell_{\text{OO}, T_i}(Y_i, T_i) = \frac{1}{n_t}\sum_{i:T_i = t}L_i(T_i)$. This is possible in standard causal inference, but not in the latent outcome setting.

Finally, we introduce two ablations of the novel Impute and Stabilize algorithm to identify the relative benefits of each component. In the Impute-only ablation we perform matrix decomposition on the observed data $\mathbf Y$ to learn $\ell_{\text{I}, T_i}(Y_i, T_i)$ for observed treatment $T_i$, we perform imputation as before, and finally use an NNLM on the imputed $\tilde{\mathbf Y}_{1-\mathbf T}$ to learn $\ell_{\text{I}, 1-T_i}(Y_i, T_i)$ for unobserved treatment $1 - T_i$. In the Stabilize-only ablation, we perform matrix decomposition on the untreated subset of observed data $\mathbf Y_{\{i:T_i = 0\}}$ to learn $\hat {\boldsymbol \lambda}$ and $\ell_{\text{S}, 0}(Y_i, T_i = 0)$ and NNLM on the treated subset $\mathbf Y_{\{i:T_i = 1\}}$ to learn $\ell_{\text{S}, 1}(Y_i, T_i = 1)$.

Table \ref{tab:algorithms} summarizes all algorithms in term of required preprocessing, the data used for NMF and NNLM steps, and the form of the final ATE estimator. Detailed pseudocode for each algorithm is provided in Appendix \ref{algorithms}. For all cases of NMF we use the R package \texttt{NMF} with the correct rank, 5 runs, and the {\tt "brunet"} algorithm to minimize KL-divergence (equivalent to maximizing Poisson likelihood) \citep{nmf_R}. After performing NMF, we rescale $\boldsymbol \lambda$ and $\mathbf L$ such that columns of the $\boldsymbol \lambda$ sum to 1. This allows causal effects to be interpreted on the original scale of the data in $\mathbf Y$. For all cases of NNLM, we use our own implementation of gradient descent to minimize KL-divergence with a fixed factor matrix, adopting the standard multiplicative updates from \citet{lee1999learning}.

\subsection{Uncertainty quantification via bootstrapping}

To quantify uncertainty in ATE estimates obtained from any of our NMF-based algorithms, we implement a bootstrap procedure. Specifically, we generate $B$ bootstrap replicates of the dataset and re-run the full estimation pipeline on each replicate.
Because each bootstrap replicate may  return a different ordering of factors, we perform post-hoc alignment of the $\hat{\boldsymbol{\lambda}}^{(b)}$ matrix before estimating $\hat{\boldsymbol\psi}_{\mathbf{L}}^{(b)}$. In simulation settings, this alignment is performed relative to the true reference matrix. In real-data applications without a known reference, we adopt an iterative consensus alignment procedure. We fix the first replicate as the reference, then sequentially align each subsequent replicates to the element-wise mean of the previously aligned matrices. Alignments use the Hungarian algorithm \citep{kuhn1955hungarian} on the negative column-wise cosine similarity matrix to maximize total aligned similarity.

After alignment, we compute the element-wise average of the aligned signature matrices:
\begin{equation*}
  \hat{\boldsymbol{\lambda}} = \frac{1}{B} \sum_{b=1}^{B} \hat{\boldsymbol{\lambda}}^{(b)}.
\end{equation*}

This is a stable consensus estimate of the signature matrix and provides the context in which to interpret bootstrapped ATE estimates. In applications, this consensus matrix may be further aligned to a reference matrix from the literature, such as the COSMIC reference for mutational signatures.

The final bootstrapped estimate of the ATE, $\hat{\boldsymbol\psi}_{\mathbf{L}}$, is computed as the element-wise average over bootstrap replicates, and $95\%$ confidence intervals as element-wise quantiles of the empirical distribution of $\hat{\boldsymbol\psi}_{\mathbf{L}}^{(b)}$, denoted $F^{-1}_{\hat{\boldsymbol\psi}_{\mathbf{L}}^{(b)}}(\cdot)$:
\begin{align*}
  \hat{\boldsymbol\psi}_{\mathbf{L}} = \frac{1}{B} \sum_{b=1}^{B} \hat{\boldsymbol\psi}_{\mathbf{L}}^{(b)}, \quad\quad
  \text{CI}_{95\%} = \left[ F^{-1}_{\hat{\boldsymbol\psi}_{\mathbf{L}}^{(b)}}(0.025),\ 
F^{-1}_{\hat{\boldsymbol\psi}_{\mathbf{L}}^{(b)}}(0.975) \right].
\end{align*}

Although consensus learned latent outcomes $\ell_{A, t}(Y_i, T_i)$ are not necessary to estimate or interpret the bootstrapped ATE, a user may need it for downstream analysis or inspection. Since the bootstrapped estimates of $\ell_{A, t}(Y_i, T_i)^{(b)}$ vary in the order and inclusion of subjects, they cannot be averaged element-wise. To recover a compatible contributions matrix, we solve an NNLM with fixed $\hat{\boldsymbol{\lambda}}$.

\section{Theoretical guarantees}

All algorithms rely on either a difference of means (DM) or a mean of individual treatment effects (mITE) estimator. The differences between the algorithms depend on two things: (1) whether estimation is performed using learned latent outcomes from a subset of observations, all observations, or all observations along with imputations, and (2) what set of data the factor model is learned with (NMF) versus a potential set of data where factors are treated as fixed (NNLM). Regardless of these specifications, the learned latent outcomes plugged into DM or mITE are estimated with NMF or NNLM.

We begin by establishing consistency results for DM and mITE estimators (Theorems~\ref{theorem:consistency_dm} and \ref{theorem:consistency_imp}). These results show that both estimators are consistent assuming that NMF and NNLM provide consistent estimates of the latent outcome and that the imputation mechanism provides consistent estimates of imputed values. Next, we show that under a set of assumptions, NMF via gradient descent on KL Divergence yields consistent estimates of the latent outcome, a conclusion that extends to the Poisson-likelihood NNLS as a special case of NMF 
(Theorem \ref{theorem:consistency_nmf}). Together,  these theorems establish the consistency of all algorithms' ATE estimates under the following assumptions:
\begin{enumerate}
  \item The distributional assumption $\mathbf Y\sim \text{Poisson}(\boldsymbol\lambda \mathbf L)$ is correctly specified.
  \item The latent dimension $K$ is correctly specified.
  \item NMF is identifiable up to the equivalence class under factor permutation and matrix scaling.
  \item Gradient descent converges to the true global optima.
  \item The imputation mechanism is consistent (mITE only).
\end{enumerate}

While these assumptions are strict and rarely fully satisfied in practice, we include these consistency results to provide theoretical reassurance under idealized conditions. A full exploration of these assumptions lies beyond the scope of this work. Notably, assumptions 1-4 are universal problems with NMF-learned latent outcomes, not specific to any of the baseline or proposed algorithms. For further discussion of identifiability and uniqueness in NMF, we refer readers to \citet{donoho2003does, laurberg2008theorems}, and \citet{huang2013non}.

\vspace{3mm}
\begin{theorem}[Consistency of difference of means estimator for ATE] \label{theorem:consistency_dm}
Let $\mathbf Y\in \mathbb R_{\ge 0}^{D\times N}$ come from a true decomposition $\mathbb E[\mathbf Y] = \boldsymbol \lambda^0 \mathbf L^0$. Assume $\hat{\boldsymbol\lambda}$ and $\hat L_{i, T_i}$ are estimated with a consistent (up to equivalence) estimator of the matrix factorization applied to observed $\mathbf{Y}(\mathbf T)$, where estimates are rescaled and permuted to match the scale and order of $\boldsymbol \lambda^0$, such that $\hat{L}_{i, T_i} \xrightarrow{p} L^0_i(T_i)
$ for all $i$, uniformly over $i$. Then, the ATE estimator
\begin{align*}
  \hat{\boldsymbol \psi}_{L,\text{DM}} &:= \frac{1}{N_1}\sum_{i: T_i = 1}\hat L_{i,T_i} - \frac{1}{N_0}\sum_{i: T_i = 0}\hat L_{i,T_i}
\end{align*}
is consistent for the true average treatment effect on the latent outcome $\boldsymbol \psi_L := \mathbb{E}[L^0(1) - L^0(0)]$ as sample size $N \to \infty$.
\end{theorem} 
\vspace{3mm}

\begin{theorem}[Consistency of mean individual treatment effect estimator for ATE] \label{theorem:consistency_imp}
Let $\mathbf Y\in \mathbb R_{\ge 0}^{D\times N}$ come from a true decomposition $\mathbb E[\mathbf Y] = \boldsymbol \lambda^0 \mathbf L^0$. Assume a consistent imputation mechanism for $\tilde{\mathbf{Y}}_{i, 1-T_i}$ such that $\tilde{{Y}}_{i, 1-T_i} \xrightarrow{p} \mathbb{E}[{Y}_i(1-T_i)]$. Also assume $\hat{\boldsymbol\lambda}$ and $\hat{L}_{i,t}$ are estimated with a consistent (up to equivalence) estimator of the matrix factorization applied to $\tilde{\mathbf{Y}}_{\mathbf t}$ (observed, imputed, or a combination), where estimates are rescaled and permuted to match the scale and order of $\boldsymbol \lambda^0$, such that $\hat{L}_{i, t} \xrightarrow{p} L^0_i(t)
$ for all $i$, uniformly over $i$. Then, the ATE estimator
\begin{align*}
  \hat{\boldsymbol \psi}_{L,\text{mITE}} &:= \frac{1}{N} \sum_{i=1}^N \left( \hat{L}_{i,1} - \hat{L}_{i,0} \right)
\end{align*}
is consistent for the true average treatment effect on the latent outcome $\boldsymbol \psi_L := \mathbb{E}[L^0(1) - L^0(0)]$ as sample size $N \to \infty$.
\end{theorem}
\vspace{3mm}

\begin{theorem}[Consistency of NMF via KL Divergence]\label{theorem:consistency_nmf}
  Let $\mathbf Y\in \mathbb R_{\ge 0}^{D\times N}$ come from a true decomposition $\mathbb E[\mathbf Y] = \boldsymbol \lambda^0 \mathbf L^0$ such that $\boldsymbol \lambda^0, \mathbf L^0 = \arg\min_{\boldsymbol\lambda, \mathbf L}\mathcal L(\boldsymbol\lambda, \mathbf L)$ with $\mathcal L$ as population KL divergence. Assuming independent columns, latent dimension $K$ is correctly specified, NMF is identifiable up to the equivalence class under permutation and scaling, and gradient descent converges to the true global minimum, $\hat{\boldsymbol \lambda}$ and $\hat{\mathbf L}$ estimated via gradient descent to minimize KL-divergence converge to their true values as sample size $N \to \infty$ (up to an equivalence class under permutation and scaling). (Proof in Appendix \ref{appendix:theory}.)
\end{theorem} 

If assumptions hold, Theorem \ref{theorem:consistency_nmf} further implies that learning-induced interference vanishes asymptotically and thus is fundamentally a finite sample problem. If $\hat{\boldsymbol \lambda}$ is consistent for the true $\boldsymbol \lambda^0$ regardless of the proportion treated in a given sample, then the decomposition is not dependent on treatment assignment mechanism as $N \to \infty$.

\section{Simulation studies} \label{sec:simulations}

\subsection{Data generation}

Simulation studies are developed in the context of mutational signatures analysis and are based on a real breast adenocarcinoma cancer dataset used in our data application (see Section \ref{sec:dataset}). We begin by fixing the latent dimension at $K = 5$ and performing NMF on all observations for a rough view of mutational signatures. These signatures are then aligned with the COSMIC v3.3.1 SBS reference signatures \citep{tate2019cosmic}. The identified reference signatures are: SBS2, SBS3, SBS6, SBS13, and SBS18. 
Using these COSMIC reference signatures as ground truth factors $\boldsymbol \lambda$, we perform nonnegative least squares (NNLS) on the \textit{untreated} samples to estimate weights. These weights are our raw sampling distribution $p(L)$ (Appendix Figure \ref{fig:sim_dist_L}). We note that there are outliers in this sampling distribution, particularly for SBS2 and SBS13. The sampling distribution for SBS3 is right skewed with a much longer tail than the other signatures.

The true ATE is set to $2000$ mutations for SBS3 and 0 for all other signatures. To simulate a single individual, we draw a full vector from $p(L)$, preserving a realistic correlation structure among signature contributions. Gaussian noise centered at zero is added to this sampled vector to generate $L(0)$. Gaussian noise centered at the ATE is added to the sampled vector to generate $L(1)$. Any negative values in either $L(0)$ or $L(1)$ are truncated to zero. Treatment is assigned using a Bernoulli distribution with 20\% probability of $T = 1$. Finally, observed data counterfactuals $Y(t)$ are generated from a Poisson distribution with rate $\boldsymbol \lambda L(t)$. We simulate 100 datasets, each consisting of 100 individuals.
Further details and pseudocode for setting simulation parameters and generating datasets can be found in Appendices \ref{appendix:sim_params} and \ref{appendix:sim_data}, respectively.

\subsection{Estimating indirect effects in simulations}

To compute the individual average learned latent outcomes under a given treatment assignment strategy $\pi$, $\bar{\ell}_{A, i}(t|\pi)$ (IALLOs, defined in Section \ref{sec:indirect_effects}), we generate $R$ realizations of resampled treatments $\mathbf T$ according to $\pi$. This is in the finite-sample regime, holding all latent potential outcomes as constant and only resampling treatment. We define $\ell_{A, t}^{\mathbf T^r}$ as a latent outcome model using algorithm $A$ and trained on data $\{\mathbf T^r, \mathbf Y(\mathbf T^r)\}$ generated from treatment program $\mathbf T^r$. The algorithm $A$ is run on each realization to learn latent outcomes $\ell^{\mathbf T^r}_{A, T^r_i}(Y_i, T^r_i)$. Treatment program $\mathbf T^{r, \sim i}$ is generated by flipping the $i^{th}$ treatment $T^r_i$. One realization $r$ and one individual $i$ at a time, algorithm $A$ is re-trained on data $\{\mathbf T^{r, \sim i}, \mathbf Y(\mathbf T^{r, \sim i})\}$ to learn latent outcome $\ell^{\mathbf T^{r\sim i}}_{A, 1-T^{r}_{i}}(Y_i, 1-T^{r}_i)$. Finally, for each individual $i$ and each treatment assignment level $t$, learned latent outcomes are averaged to estimate $\bar{\ell}_{A, i}(t|\pi)$:
\begin{align*}
  \bar{\ell}_{A, i}(t|\pi) &= \mathbb E_{\mathbf{T}_{-i} \sim \pi}\left[ \ell_{A, t}(Y_i, t) | T_i = t, \mathbf T_{-i}\right]\\
  & \approx \frac{1}{R}\sum_{r = 1}^R\left(
    \ell^{\mathbf T^r}_{A, T^r_i}(Y_i, T^r_i)\cdot I(T_i^r = t) +
    \ell^{\mathbf T^{r\sim i}}_{A, 1-T^{r}_{i}}(Y_i, 1-T^{r}_i)\cdot I(1-T^{r}_i = t)
  \right); \quad \forall r, \mathbf T^r \sim \pi . 
\end{align*}
In the case of the Random Split algorithm, to compute the IALLO for individual $i$, we ensure that $i$ is in the held out data ($i \ne S$) so that its latent outcome is learned. In our simulation studies, we use $R = 20$ realizations. 

We do not use bootstrapped estimates here because indirect effects are intended to capture perturbations due to changing treatment, whereas bootstrap resampling captures variability due to sample selection.

\subsection{Results: learning-induced interference}

\begin{figure}
  \centering
  \includegraphics[height=0.8\linewidth]{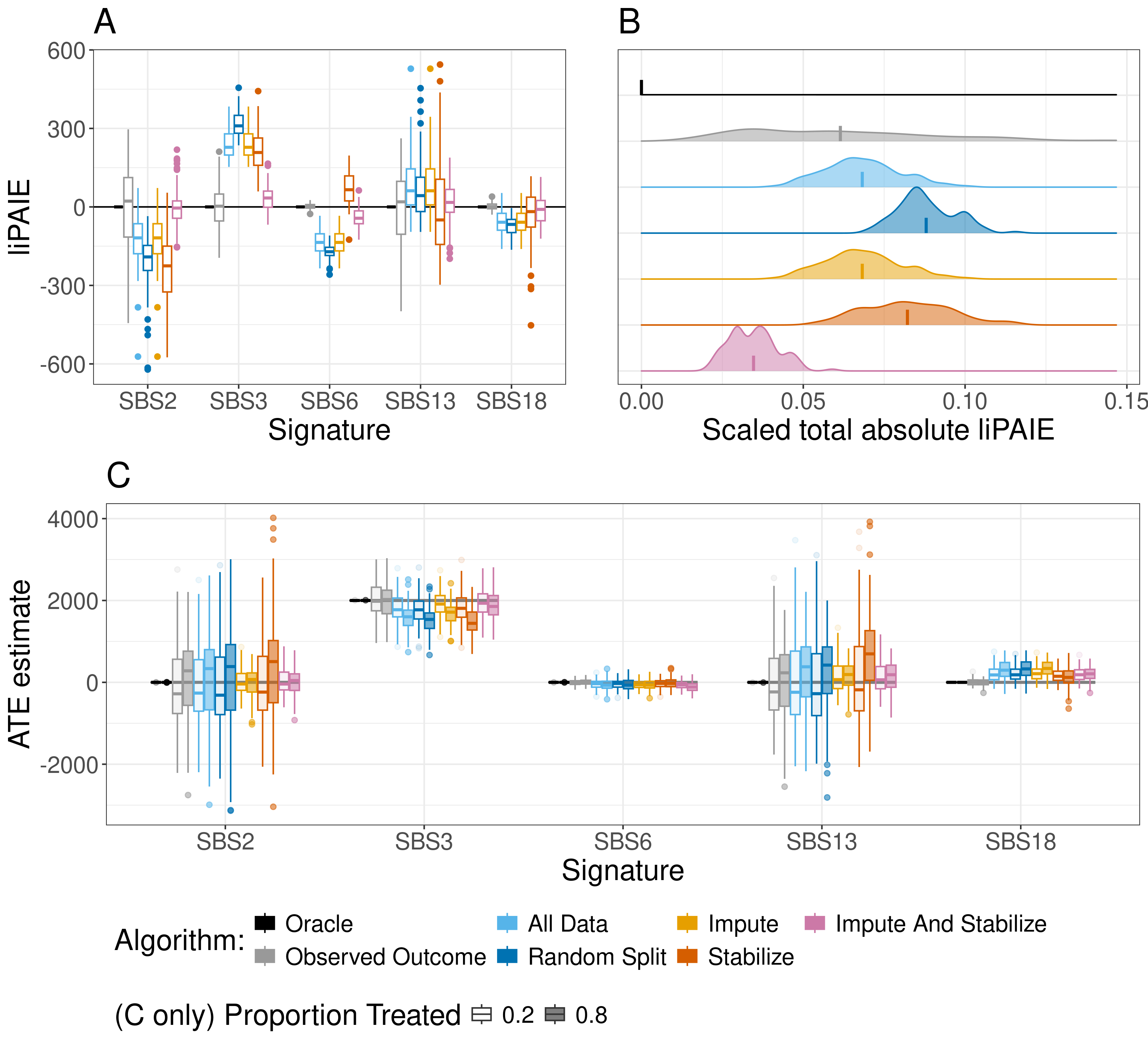}
  \caption{\textbf{Simulation results: indirect effects} across 100 simulated datasets of 100 individuals each. \textbf{A) Learning-induced population average indirect effects (liPAIEs)} for 5 cancer mutational signatures. This represents the expected change in a single dimension of an untreated individual's learned latent outcome (i.e., number of mutations attributed to the given signature) when other subjects change from 20\% treated to 80\% treated. Without learning-induced interference, these values will be centered at 0. \textbf{B) Sum of absolute liPAIEs per sample}, rescaled by two times the number of mutations per sample (because any change is counted twice: by liIAIE of its old and new signature attribution). This represents the proportion of mutations, per individual, whose attribution changes due to the shift of other subjects from 20\% treated to 80\% treated. Mean values per algorithm are marked with vertical ticks. \textbf{C) Bootstrapped mean ATE estimates} with either 20\% treated (filled in white) or 80\% treated (filled in with color). Under no learning-induced interference, we expect the same ATE estimates regardless of the proportion treated.}
  \label{fig:simulation_indirect}
\end{figure}

The full Impute and Stabilize algorithm substantially reduces learning-induced indirect effects compared to baselines and ablations. Specifically, it achieves learning-induced population average indirect effects (liPAIEs) centered at zero (Figure \ref{fig:simulation_indirect}A), reduces the scaled total absolute liPAIEs by a factor of at least two (Figure \ref{fig:simulation_indirect}B), and produces ATE estimates appearing unchanged between 20\% and 80\% of subjects treated (Figure \ref{fig:simulation_indirect}C).

Intriguingly, we observe that ATE estimates for SBS2 and SBS13, the two signatures with outliers in the true latent outcome distribution, vary substantially between 20\% and 80\% treated for the Observed Outcome algorithm (Figure \ref{fig:simulation_indirect}C, grey). This illustrates that even a standard causal inference outcome model on non-latent outcomes---here, the group mean---can exhibit learning-induced interference, especially in settings with outliers that become heavily influential observations. Other methods that show ATE variation for SBS2 and SBS13 (all but Oracle and Impute and Stabilize) also exhibit similar instability for other signatures. This underscores our key motivation for this work: learning-induced interference is magnified by the necessity to model such a complex and unsupervised latent structure.

The Stabilize ablation seems to improve upon the baselines in learning-induced population average indirect effects (liPAIEs) for all signatures excluding SBS2, but also displays the most variability (Figure \ref{fig:simulation_indirect}A). This variability makes sense as only a subset of the data is used in matrix decomposition, similar to Random Split. The scaled total absolute liPAIEs for the Stabilize ablation are improved from Random Split, but worse than All Data or Impute due to its shortcomings in SBS2 and large variability (Figure \ref{fig:simulation_indirect}B). These results suggest that stabilization drives the Impute and Stabilize algorithm's reduction in learning-induced interference, but it is clear that stabilization must be combined with imputation to achieve full benefits.

\subsection{Results: bias and efficiency}

\begin{figure}
  \centering
  \includegraphics[height=0.7\linewidth]{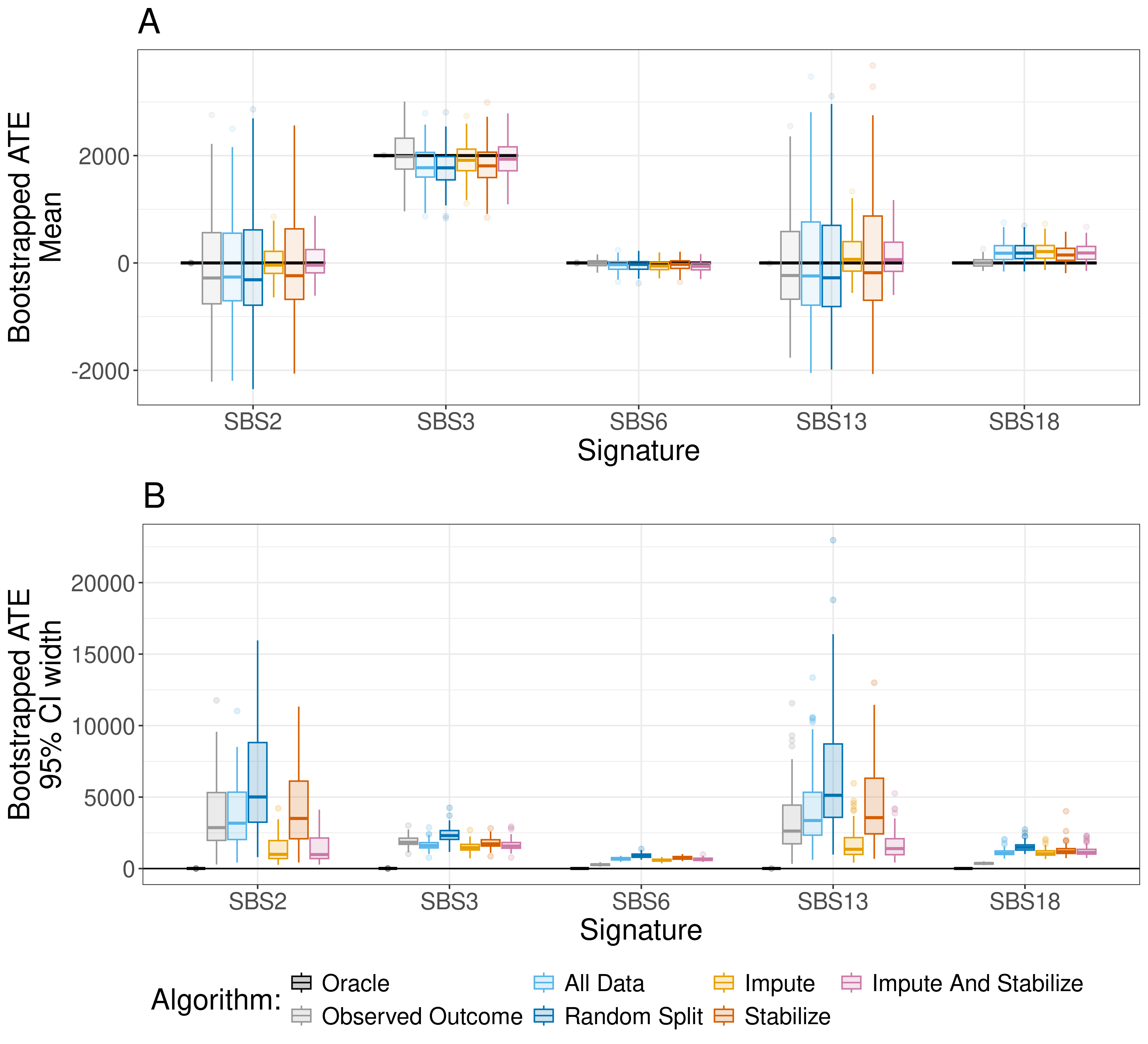}
  \caption{\textbf{Simulation results: average treatment effects (ATEs).} \textbf{A) Bootstrapped mean ATEs} across 100 simulated datasets. Black lines indicate ground truth ATE. \textbf{B) Bootstrapped 95\% confidence interval widths} across 100 simulated datasets with black lines at zero.}
  \label{fig:simulation_direct}
\end{figure}

\renewcommand{\arraystretch}{1.5}
\begin{table}[t]
  \centering
  \begin{tabular}{|c||c|c|c|c|c|}
    \hline
    \textbf{Algorithm} & \textbf{SBS2} & \textbf{SBS3} & \textbf{SBS6} & \textbf{SBS13} & \textbf{SBS18}\\
    \hline
    Oracle & 0.91 & 0.92 & 0.95 & 0.94 & 0.94\\
    Observed Outcome & 0.79 & 0.96 & 0.94 & 0.80 & 0.96\\
    \hline
    All Data & 0.86 & 0.94 & 0.98 & 0.86 & 0.98\\
    Random Split & 0.95 & 0.95 & 0.99 & 0.98 & 1.00\\
    \hline
    Impute & 0.91 & 0.95 & 0.98 & 0.96 & 0.99\\
    Stabilize & 0.86 & 0.94 & 0.99 & 0.86 & 1.00\\
    Impute And Stabilize & 0.91 & 0.96 & 0.98 & 0.94 & 1.00\\
    \hline
  \end{tabular}
  \caption{\textbf{Coverage}, or proportion of datasets where the 95\% confidence interval includes the true ATE for each signature, across 100 simulated datasets. Coverage < 0.95 indicates undercoverage and may imply unreliable estimate of the ATE. Coverage > 0.95 means the confidence intervals are conservative, or wider than they need to be. These values are only precise to the decimals reported, as only 100 simulated datasets were used.}
  \label{tab:coverage}
\end{table}

Both our Impute and Stabilize algorithm and our Impute-only ablation show reduced bias in the bootstrapped ATE mean compared to the baselines, and also show less variability around such estimates (Figure \ref{fig:simulation_direct}A). Moreover, especially for the two signatures with outliers in the data generating distribution (SBS2 and SBS13), these two algorithms even outperform the Observed Outcome algorithm, a hypothetical world in which latent outcomes are \textit{directly observed}.
This highlights that, despite the challenges of estimating the factor model \textit{de novo}, imputation yields great benefits by better accounting for sample-to-sample variability and providing robustness to outliers in the latent outcome distribution.

These two algorithms also show much narrower confidence intervals than any baseline, again in particular for the two signatures with outliers in the latent outcome distribution (Figure \ref{fig:simulation_direct}B). The coverage values reported in Table \ref{tab:coverage} assure us that despite the reduced interval width, coverage remains at or above the Oracle level. For other signatures, confidence interval widths are comparable to most other algorithms, and still narrower than Random Split and Stabilize. This is expected, as Random Split and Stabilize both use a subset of the data to fit the factor model, increasing variability.

\section{Application: effect of germline BRCA mutations on cancer mutational signatures in early-onset breast adenocarcinoma}

\begin{figure}
  \centering
  \includegraphics[width=0.9\linewidth]{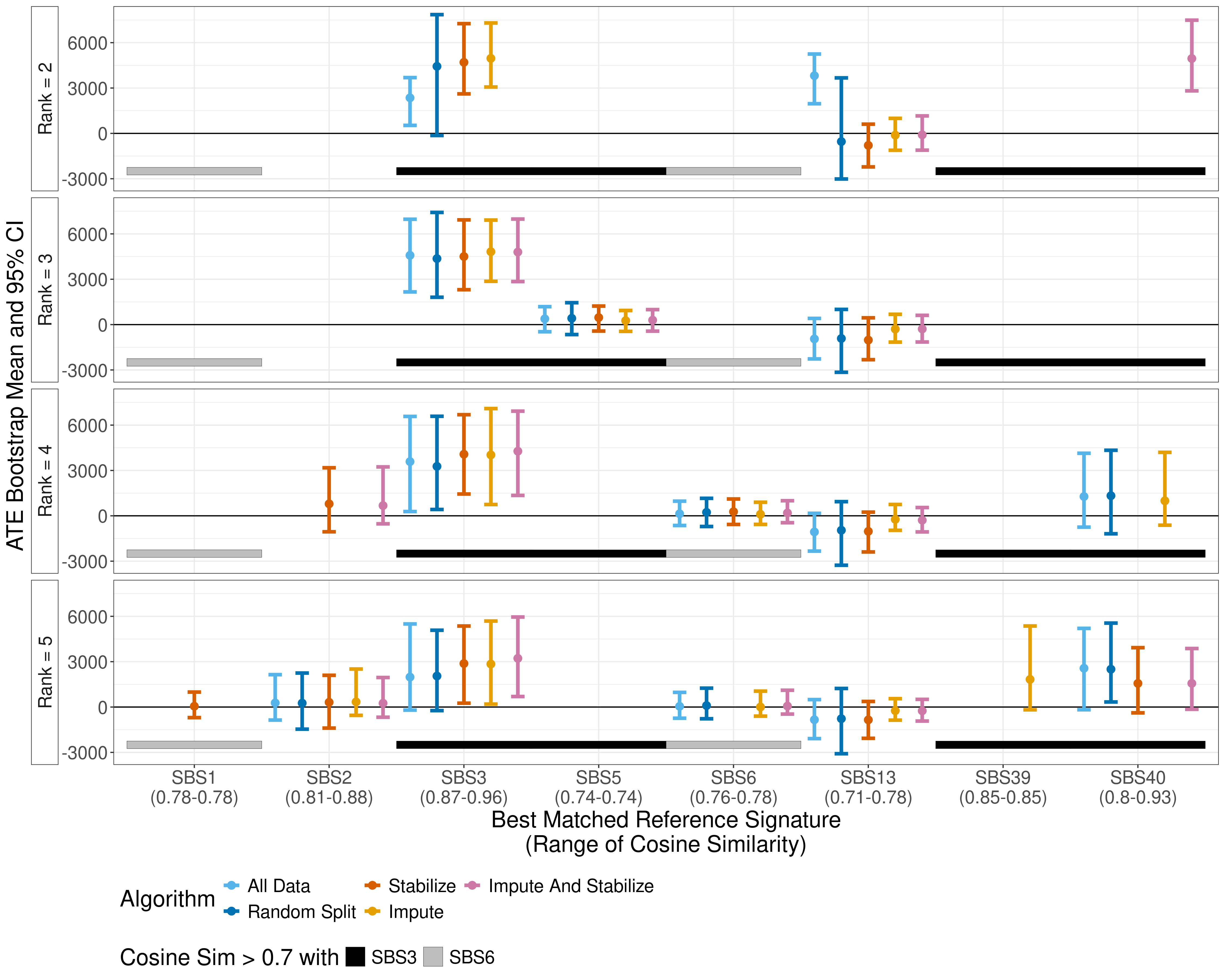}
  \caption{\textbf{Early onset breast adenocarcinoma results.} Bootstrapped means and 95\% confidence intervals across baseline, ablation, and novel algorithms. Results shown for ranks 2-5, top to bottom. Black rectangles along the x-axis indicate COSMIC reference signatures with cosine similarity to SBS3 $> 0.7$. Grey rectangles along the x-axis indicate COSMIC reference signatures with cosine similarity to SBS6 $> 0.7$. The x-axis labels report minimum and maximum cosine similarity between estimated and reference for each signature, across ranks and algorithms. Individual cosine similarity results can be found in Appendix Figure \ref{fig:application_sim}.}
  \label{fig:application_res}
\end{figure}

\subsection{Background and causal question}
Germline genetic variants, or those present in an individual from birth, provide a natural treatment variable for causal inference. There is a precedent for assuming near-randomization of these variables, as Mendelian Randomization methods in causal inference use germline variants for instrumental variables \citep{davey2003mendelian}. Unlike environmental or behavioral exposures, germline variants are not directly influenced by covariates, and they unambiguously precede the latent outcome of mutational signature contributions. A prominent example in mutational signatures analysis is the relationship between BRCA1/2 germline mutations and the COSMIC reference signature SBS3, especially in breast and ovarian cancers. SBS3 reflects the same defects in homologous recombination repair that pathogenic BRCA mutations cause \citep{nik2012mutational, alexandrov2013signatures, nik2016landscape, chen2022brca1}. Because similar deficiencies can arise through somatic BRCA mutations acquired later in life, this link between germline BRCA status and SBS3 is particularly strong in early-onset breast adenocarcinoma \citep{andrikopoulou2022mutational}. In this section, we estimate the causal effect of carrying at least one pathogenic germline mutation in the BRCA1 and/or BRCA2 genes on the number of mutations attributed to mutational signatures in early-onset breast adenocarcinoma.

\subsection{Dataset}\label{sec:dataset} 

We accessed the publicly available mutational counts data for the whole genome sequencing (WGS) 96-alphabet mutation classification from International Cancer Genome Consortium's (ICGC) Accelerating Research in Genomic Oncology (ARGO) data portal [\citeauthor{zhang2019international}, \citeyear{zhang2019international}, \href{https://docs.icgc-argo.org/docs/data-access/icgc-25k-data#open-release-data---object-bucket-details}{access instructions link}]. We were also granted access to the private ICGC ARGO data and accessed the legacy Pan Cancer Analysis of Whole Genomes (PCAWG) \citep{icgc2020pan} data through their SFTP server. The \texttt{germline\_variations} subdirectory contains the results of germline mutation calling as described in \citet{icgc2020pan}, the {\tt clinical\_and\_histology} subdirectory contains age information, and the \texttt{donors\_and\_biospecimens} subdirectory contains a mapping between IDs used for somatic (tumor) mutational counts and IDs used for germline (normal tissue) mutational calling.

We restricted analysis to subjects with breast adenocarcinoma histology labels. While somatic mutation data is widely available, germline variant calling is only available for the 111 non-US subjects. Of the subjects available in both data modalities, 
we further subset to focus on early-onset breast adenocarcinomas, defined by an age of diagnosis younger than 45 years \citep{Clinic_2025}, yielding 27 individuals. Note that this age subsetting was not applied during simulation study design.

All recorded germline variants were subset to those in genomic regions corresponding to BRCA1 and BRCA2 genes. Variants were annotated using ANNOVAR \citep{wang38annovar} and subset to 65 variants annotated as "pathogenic" or "likely pathogenic". We then identified 3 individuals with at least one pathogenic germline mutation in the BRCA1 and/or BRCA2 genes. The presence of at least one such mutation is the binary treatment in our causal question.

Although we cannot release this dataset publicly, all data processing code are available on GitHub at \href{https://github.com/jennalandy/causalLFO_PAPER}{\mbox{jennalandy/causalLFO\_PAPER}},  for use by researchers with approval to access the private ICGC ARGO data.

Latent dimension selection and its impact of that choice on consistency or interference has not been the focus of this paper. For the data application, we instead report results for a range of latent ranks determined ``reasonable''---based on a standard survey of NMF metrics and avoiding duplicated signatures---and report all results. NMF was run on this final dataset for ranks between 1 and 15, and latent ranks of 2-5 mutational signatures were determined to be a reasonable range (Appendix Figure \ref{fig:rank_sel_nmf_BA_sub45}).

\subsection{Results}

The causal effect of carrying at least one pathogenic germline mutation in the BRCA1 or BRCA2 genes on the number of mutations attributed to mutational signatures is nearly always significantly positive for the signature most closely matching COSMIC reference SBS3, especially as estimated by our novel algorithm (Impute and Stabilize) or any of its ablations (Figure \ref{fig:application_res}). A notable exception is with a rank of 2, where Impute and Stabilize more closely estimates SBS40 in place of SBS3. However, these signatures have a cosine similarity of 0.88 so may be reasonably interchanged. The magnitude of the causal effect on SBS3 contributions decreases as rank increases above 4, likely due to the incorporation of other signatures that have high cosine similarity to SBS3: SBS5, SBS39, or SBS40. For the most part, the same signatures---or at least similar signatures in terms of cosine similarity---are chosen by all algorithms for each rank.

Confidence interval widths show that Impute and Stabilize is more efficient than the baselines, particularly for SBS3 and SBS13. For these same signatures, there is a noticeable difference in mean values where the means reported by Impute and Stabilize are higher than the baselines by up to 1000 mutations. This pattern suggests that learning-induced interference is affecting the ATEs of baseline algorithms as seen in our simulations. In terms of statistical decision-making, the Random Split and All Data algorithms yield different significance decisions depending on the rank, while the Impute and Stabilize algorithm has robust conclusions.

While the association between germline BRCA1 or BRCA2 mutations and signature SBS3 is well-established in early-onset breast adenocarcinoma  \citep{nik2012mutational, alexandrov2013signatures, nik2016landscape, chen2022brca1}, our contribution is to formalize this relationship within a causal inference framework. Our method allows for improved ATE estimates with more efficiency and less impact from learning-induced interference. From the results of our simulation study, it is clear that the estimates yielded by the Impute and Stabilize algorithm are more reliable than those from either baseline.

\section{Discussion and future work}

In this paper, we formalized the difference between interference in a data generating process and learning-induced interference, and we introduced a quantification of the latter using indirect effects. We proposed a new algorithm to estimate causal ATE on Poisson likelihood NMF-learned latent outcomes that significantly reduces learning-induced interference while improving estimation efficiency. These benefits were demonstrated in simulation studies, and our real-data application provides a realistic, hypothesis-driven example of this algorithm in practice. To the best of our knowledge, this is the first work to formally address causal inference on latent outcomes derived from NMF. While our simulation studies and data application are in the context of cancer mutational signatures, the proposed algorithm is generalizable to any latent outcomes learned via Poisson NMF, and the concepts we introduced may be more broadly applied to latent outcomes obtained by other factor models.

We emphasize that the Impute and Stabilize algorithm does not \textit{fully resolve} the issue of learning-induced interference, though it does make a substantial step in that direction. As discussed in Section \ref{sec:indirect_effects}, learning-induced indirect effects at zero do not necessarily imply the absence of learning-induced interference, only that such interference has no effect on mean learned latent outcomes, and thus no effect on the ATE. Although the Impute and Stabilize algorithm has clearly reduced the magnitude of learning-induced indirect effects, they are not exactly zero, and can still impact our ATE estimates. Further, residual learning-induced interference not captured by indirect effects, such as dependencies affecting the variance of learned latent outcomes, may impact our estimated confidence intervals.

We also acknowledge potential limitations of this approach in sparse settings, or generally in settings where the magnitude of the causal effect on observed data $Y$ is greater than the scale of $Y$ itself. In such cases, imputation may yield negative imputed values on the square-root scale, requiring additional modifications.

Importantly, if a factor appears in the treated condition only, it will not be captured by our stabilization approaches where the factor model is fit on untreated data alone. Additionally, as explored in the data application, this algorithm is still sensitive to the choice of latent rank. In practice, we recommend investigating results across a range of plausible ranks, as done here. Seeing results mirrored across varying levels of granularity helps reassure us of a meaningful signal. In our case, results line up in terms of the significance decision but not necessarily in terms of magnitude, so point estimates of ATEs may be difficult to interpret or trust.

We reiterate that even outcome models used in standard causal inference, such as those used in g-computation or AIPW, can exhibit learning-induced interference, particularly in the presence of outliers or influential observations. These effects are often attributed to small-sample variability and not explicitly modeled. However, our framework of learning-induced interference and associated metrics may prove useful to formally quantify this issue in other areas of causal inference. For instance, learning-induced population average indirect effects could be used to quantify the extent to which strategies like cross-fitting mitigate learning-induced interference.

Our data application focused on germline mutations as near-randomized treatments, though this work could apply to randomized clinical trials or other designs where randomization is guaranteed. Future extensions of the Impute and Stabilize algorithm could incorporate covariates for applications in observational studies where confounding adjustment is required.

We see many directions for future work to further improve the Impute and Stabilize algorithm. First, covariates could easily be incorporated at the imputation stage and at the ATE estimation stage, though incorporating covariates in NMF may prove more difficult. Second, with method-specific choices for the imputation strategy, the algorithm could be extended to other classical factor models, such as factor analysis, or adapted to deep representation learning methods like language models and graph neural networks. Finally, Bayesian NMF could be utilized instead of bootstrapping as an alternative way to quantify uncertainty.

This work provides a formal foundation for causal inference on latent outcomes, addresses a critical gap in handling learning-induced interference, and introduces a practical and effective algorithm that advances both theory and application in this emerging area.

\section*{Acknowledgments}
The authors gratefully acknowledge that this work was supported by the NIH-NCI grant 5R01 CA262710-03 and NSF-DMS grant 2113707.

\bibliographystyle{unsrtnat}
\bibliography{references}

\newpage
\appendix
\counterwithin{figure}{section}
\counterwithin{table}{section}
\pagenumbering{arabic}
  \setcounter{page}{1}

\begin{center}
  \large \textbf{Appendix}
\end{center}
\vspace{-0.5cm}

\label{sec:appendix}

\section{Algorithm definitions} \label{algorithms}

\begin{algorithm}[H]
\caption{All Data}
\textbf{Input:} Observed data $\mathbf Y \in \mathbb R_{\ge 0}^{D\times N}$, treatment vector $\mathbf T = [T_1, T_2, \dots, T_N]$
\begin{algorithmic}[1]
\State \textbf{Fit factor model}: 
\State \quad Perform NMF on full $\mathbf Y$ to estimate $\hat{\boldsymbol\lambda}$ and $\hat {\mathbf L}$
\State \quad Normalize $\hat{\boldsymbol\lambda}$ and $\hat{\mathbf L}$ so columns of $\hat{\boldsymbol\lambda}$ sum to 1
\State \quad $\ell_{\text{AD}, T_i}(Y_i, T_i) = \hat L_i$
\State \textbf{Estimate causal effect}:
\State \quad Compute $\hat{\boldsymbol \psi}_{L, \text{AD}} = \frac{1}{N_1}\sum_{i:T_i = 1} \ell_{\text{AD}, T_i}(Y_i, T_i) - \frac{1}{N_0}\sum_{i:T_i = 0} \ell_{\text{AD}, T_i}(Y_i, T_i)$
\end{algorithmic}
\end{algorithm}

\begin{algorithm}[H]
\caption{Random Split}
\textbf{Input:} Observed data $\mathbf Y \in \mathbb R_{\ge 0}^{D\times N}$, treatment vector $\mathbf T = [T_1, T_2, \dots, T_N]$, split proportion $p$ (default $p = 1/2$)
\begin{algorithmic}[1]
\State \textbf{Preprocessing}: 
\State \quad Randomly sample indices $S \subset \{1,\dots, N\}$ such that $||S|| = \lceil N\cdot p \rceil$
\State \textbf{Fit factor model}: 
\State \quad Perform NMF on $\mathbf Y_S$ to estimate $\hat{\boldsymbol\lambda}$
\State \quad Normalize columns of $\hat{\boldsymbol\lambda}$ so they sum to 1
\State \textbf{Estimate causal effect}:
\State \quad Estimate $\hat{\mathbf L}_{/S}$ by applying nonnegative linear model to $\mathbf Y_{/S}$ with fixed $\hat{\boldsymbol\lambda}$
\State \quad $\ell_{\text{RS}, T_i}(Y_i, T_i) = \hat L_i$ for $i \notin S$
\State \quad Compute $\hat{\boldsymbol \psi}_{L, \text{RS}} = \frac{\sum_{i \notin S}\ell_{\text{RS}, T_i}(Y_i, T_i) I(T_i = 1)}{\sum_{i \notin S} I(T_i = 1)} - \frac{\sum_{i \notin S}\ell_{\text{RS}, T_i}(Y_i, T_i) I(T_i = 0)}{\sum_{i \notin S} I(T_i = 0)}$
\end{algorithmic}
\end{algorithm}

\begin{algorithm}[H]
\caption{Impute}
\textbf{Input:} Observed data $\mathbf Y \in \mathbb R_{\ge 0}^{D\times N}$, treatment vector $\mathbf T = [T_1, T_2, \dots, T_N]$
\begin{algorithmic}[1]
\State \textbf{Preprocessing}:
\State \quad Construct $\tilde{\mathbf Y}_{1-\mathbf T}$ by imputing unobserved potential outcome for each sample with $f_{\text{IMP}}$ (Algorithm \ref{alg:fimp})
\State \textbf{Fit factor model}:
\State \quad Perform NMF on observed $\mathbf Y$ to estimate $\hat{\boldsymbol\lambda}$ and $\hat{\mathbf L}_{\mathbf T}$
\State \quad Normalize $\hat{\boldsymbol\lambda}$ and $\hat{\mathbf L}_{\mathbf T}$ so columns of $\hat{\boldsymbol\lambda}$ sum to 1
\State \quad $\ell_{\text{I}, T_i}(Y_i, T_i) = \hat L_{T_i, i}$, the $i^{th}$ column of $\hat{\mathbf L}_{\mathbf T}$
\State \textbf{Estimate causal effect}:
\State \quad Estimate $\hat{\mathbf L}_{1-\mathbf T}$ by applying nonnegative linear model to $\tilde{\mathbf Y}_{1-\mathbf T}$ with fixed $\hat{\boldsymbol\lambda}$
\State \quad $\ell_{\text{I}, 1-T_i}(Y_i, T_i) = \hat L_{1-\mathbf T, i}$, the $i^{th}$ column of $\hat{\mathbf L}_{1-\mathbf T}$
\State \quad Compute $\hat{\boldsymbol \psi}_{L, \text{I}} = \frac{1}{N} \sum_{i=1}^N \left( \ell_{\text{I}, 1}(Y_i, T_i) - \ell_{\text{I}, 0}(Y_i, T_i) \right)$
\end{algorithmic}
\end{algorithm}

\begin{algorithm}[H]
\caption{Stabilize}
\textbf{Input:} Observed data $\mathbf Y \in \mathbb R_{\ge 0}^{D\times N}$, treatment vector $\mathbf T = [T_1, T_2, \dots, T_N]$
\begin{algorithmic}[1]
\State \textbf{Fit factor model}:
\State \quad Perform NMF on untreated samples $\mathbf Y_{\{i:T_i = 0\}}$ to estimate $\hat{\boldsymbol\lambda}$ and $\hat{\mathbf L}_0$
\State \quad Normalize $\hat{\boldsymbol\lambda}$ and $\hat{\mathbf L}_0$ so columns of $\hat{\boldsymbol\lambda}$ sum to 1
\State \quad $\ell_{\text{S}, 0}(Y_i, T_i) = \hat{\mathbf L}_{0,i}$ for $i$ with $T_i = 0$
\State \textbf{Estimate causal effect}:
\State \quad Estimate $\hat{\mathbf L}_1$ by applying nonnegative linear model to $\mathbf Y_{\{i:T_i = 1\}}$ with fixed $\hat{\boldsymbol\lambda}$
\State \quad $\ell_{\text{S}, 1}(Y_i, T_i) = \hat{\mathbf L}_{1,i}$ for $i$ with $T_i = 1$
\State \quad Compute $\hat{\boldsymbol \psi}_{L, \text{S}} = \frac{1}{N_1}\sum_{i:T_i = 1} \ell_{\text{S}, T_i}(Y_i, T_i) - \frac{1}{N_0}\sum_{i:T_i = 0} \ell_{\text{S}, T_i}(Y_i, T_i)$
\end{algorithmic}
\end{algorithm}

\begin{algorithm}[H]
\caption{Impute and Stabilize}
\textbf{Input:} Observed data $\mathbf Y \in \mathbb R_{\ge 0}^{D\times N}$, treatment vector $\mathbf T = [T_1, T_2, \dots, T_N]$
\begin{algorithmic}[1]
\State \textbf{Preprocessing}:
\State \quad Construct $\tilde{\mathbf Y}_{1-\mathbf T}$ by imputing unobserved potential outcome for each sample with $f_{\text{IMP}}$ (Algorithm \ref{alg:fimp})
\State \quad Create $\tilde{\mathbf Y}_0$ and $\tilde{\mathbf Y}_1$ such that $\tilde{\mathbf Y}_{t, i} = \mathbf Y_i$ if $T_i = t$ and $\tilde{\mathbf Y}_{t, i} =  \tilde{\mathbf Y}_{1-\mathbf T, i}$ if $T_i = 1-t$
\State \textbf{Fit factor model}:
\State \quad Perform NMF on $\tilde{\mathbf Y}_0$ to estimate $\hat{\boldsymbol\lambda}$ and $\hat{\mathbf L}_0$
\State \quad Normalize $\hat{\boldsymbol\lambda}$ and $\hat{\mathbf L}_0$ so columns of $\hat{\boldsymbol\lambda}$ sum to 1
\State \quad $\ell_{\text{IS}, 0}(Y_i, T_i) = \hat{\mathbf L}_{0,i}$ for all $i$
\State \textbf{Estimate causal effect}:
\State \quad Estimate $\hat{\mathbf L}_1$ by applying nonnegative linear model to $\tilde{\mathbf Y}_1$ with fixed $\hat{\boldsymbol\lambda}$
\State \quad $\ell_{\text{IS}, 1}(Y_i, T_i) = \hat{\mathbf L}_{1,i}$ for all $i$
\State \quad Compute $\hat{\boldsymbol \psi}_{L, \text{IS}} = \frac{1}{N} \sum_{i=1}^N \left(\ell_{\text{IS}, 1}(Y_i, T_i) - \ell_{\text{IS}, 0}(Y_i, T_i) \right)$
\end{algorithmic}
\end{algorithm}

\begin{algorithm}[H]
\caption{Imputation Function $f_{\text{IMP}}$} \label{alg:fimp}
\textbf{Input:} Count matrix $\mathbf Y \in \mathbb{R}_{\ge 0}^{D \times N}$, treatment vector $\mathbf T = [T_1, \dots, T_G]$
\begin{algorithmic}[1]
\State \textbf{Variance stabilization}:
\State \quad Compute $\mathbf Y^{\text{vst}} = \sqrt{\mathbf Y}$
\State \textbf{Estimate ATE on stabilized scale}:
\State \quad Compute $\hat{\boldsymbol\psi}_{\mathbf{Y}}^{\text{vst}} = \frac{1}{N_1} \sum_{i: T_i = 1} Y^{\text{vst}}_i 
  - \frac{1}{N_0} \sum_{i: T_i = 0} Y^{\text{vst}}_i$
\State \textbf{Impute counterfactuals on stabilized scale}:
\State \quad $\tilde{Y}^{\text{vst}}_{1-T_i, i} = Y^{\text{vst}}_i + (1 - T_i) \cdot \hat{\boldsymbol\psi}_{\mathbf{Y}}^{\text{vst}} - T_i \cdot \hat{\boldsymbol\psi}_{\mathbf{Y}}^{\text{vst}}$
\State \textbf{Back-transformation}:
\State \quad $\tilde{Y}_{1-T_i, i} = \left( \tilde{Y}^{\text{vst}}_{1-T_i, i} \right)^2 + \frac{1}{4} \left( 1 + \frac{1}{N_1} + \frac{1}{N_0} \right)$
\end{algorithmic}
\end{algorithm}

\newpage
\section{Theoretical guarantees}\label{appendix:theory}

\textbf{Theorem} (Consistency of NMF via KL Divergence)
  Let $\mathbf Y\in \mathbb R_{\ge 0}^{D\times N}$ come from a true decomposition $\mathbb E[\mathbf Y] = \boldsymbol \lambda^0 \mathbf L^0$ such that $\boldsymbol \lambda^0, \mathbf L^0 = \arg\min_{\boldsymbol\lambda, \mathbf L}\mathcal L(\boldsymbol\lambda, \mathbf L)$ with $\mathcal L$ as population KL divergence. Assuming independent columns, latent dimension $K$ is correctly specified, NMF is identifiable up to the equivalence class under permutation and scaling, and gradient descent converges to the true global minimum, $\hat{\boldsymbol \lambda}$ and $\hat{\mathbf L}$ estimated via gradient descent to minimize KL-divergence converge to their true values as sample size $N \to \infty$ (up to an equivalence class under permutation and scaling). 
\begin{proof} The proof for this theorem begins with the convergence of empirical loss and uses M-estimation theory to prove the consistency of loss minimizers. Assuming gradient descent converges to a global minimum, this shows that NMF estimates are consistent. We first show consistency of the factor matrix estimator $\hat{\boldsymbol \lambda}$ alone, then infer convergence of individual weight estimators $\hat L_i$.

  \textbf{Definitions}:
  \begin{itemize}
    \item Recall the matrix $\mathbf Y$ contains the observed data, with each column $Y_i$ representing a $D$-dimensional data vector for subject $i$. In expectations, we let $Y$ denote a $D$-dimensional random variable. The same applies for latent outcomes matrix $\mathbf L$, its columns $L_i$, and a $K$-dimensional random variable $L$.
    \item Loss function for a $D$-dimensional column vector $Y_i$
    \begin{align*}
        KL(Y_i \,\|\, \boldsymbol \lambda L_i) & = \sum_d\left(Y_{di} \log \frac{Y_{di}}{(\boldsymbol \lambda L_i)_{d}} - Y_{di} + (\boldsymbol \lambda L_i)_{d}\right)
    \end{align*}
    where $Y_{di}$ is the $d^{th}$ element in column vector $Y_i$.
    \item Population risk
    \begin{align*}
      \mathcal{L}(\boldsymbol \lambda, \mathbf{L}) &= \mathbb{E}_{L \in\mathbf L, Y \sim \text{Poisson}(\boldsymbol{\lambda}^0 L^0)} \left[ KL (Y \,\|\, \boldsymbol \lambda L) \right]. \\
    \end{align*}
    \item Empirical loss and estimator definition \begin{align*}
        L(\boldsymbol \lambda, \mathbf L) & =
        \sum_{i = 1}^N KL(Y_i|| \boldsymbol \lambda L_i)\\
      \hat{\boldsymbol \lambda}, \hat{\mathbf L} &= \arg \min_{\boldsymbol \lambda, \mathbf L}  L(\boldsymbol \lambda, \mathbf L).
    \end{align*}
    \item Equivalence class of NMF solutions: for any $\boldsymbol \lambda, \mathbf L $, a permutation matrix $\Pi$ and positive diagonal scaling matrix $S$ can be applied as follows
    \begin{align*}
      \boldsymbol \lambda' &= \boldsymbol \lambda S\Pi,\\ 
      \mathbf L' &= \Pi^{-1} S^{-1} \mathbf L,
    \end{align*}
    such that the same product matrix and thus the same value of KL-divergence are retained
    \begin{align*}
      \boldsymbol \lambda'\mathbf L' &= \boldsymbol \lambda \mathbf L,\\
      KL(\mathbf Y|| \boldsymbol \lambda' \mathbf L') & = KL(\mathbf Y|| \boldsymbol \lambda \mathbf L).
    \end{align*}
    This equivalence class is denoted 
    \begin{align*}
      eq(\boldsymbol \lambda, \mathbf L) = \{\boldsymbol \lambda' = \boldsymbol \lambda S\Pi, \mathbf L' = \Pi^{-1} S^{-1} \mathbf L | \text{ permutation }\Pi, \text{positive diagonal }S\}.
    \end{align*}
  \end{itemize}

  \textbf{Assumptions}:
  \begin{itemize}
    \item Latent dimension $K$ is correctly specified.
    \item The true factorization $\boldsymbol \lambda^0, \mathbf L^0$ is identifiable up to the equivalence class under permutation and scaling. That is, the equivalence class of $\boldsymbol \lambda^0, \mathbf L^0$ holds all minimizers of the population risk:
    \begin{align*}
        eq(\boldsymbol \lambda^0, \mathbf L^0) &= \arg\min_{\boldsymbol \lambda, \mathbf L}\mathbb E\left[KL( Y|| \boldsymbol \lambda L)\right].
    \end{align*}
    \item Gradient descent converges to a global minimum of KL divergence 
    \begin{align*}
       KL(\mathbf Y|| \hat{\boldsymbol \lambda} \hat{\mathbf L}) &= \min_{\boldsymbol \lambda, \mathbf L}\sum_{i = 1}^NKL( Y_i|| \boldsymbol \lambda L_i)\\
       \iff \hat{\boldsymbol \lambda}, \hat{\mathbf L} &\in \arg\min _{\boldsymbol \lambda, \mathbf L}\sum_{i = 1}^NKL( Y_i|| \boldsymbol \lambda L_i).
    \end{align*}
  \end{itemize}

  \textbf{Proof}:
  \begin{enumerate}
    \item \textbf{Uniform convergence of the empirical KL loss.} \\
    The empirical KL objective (normalized by \( N \)) converges uniformly to the population KL objective as \( N \to \infty \), due to the law of large numbers. Specifically,
    \[
    \sup_{(\boldsymbol \lambda, \mathbf L) \in \mathcal{F}} \left| \frac{1}{N} \sum_{i = 1}^N KL(Y_i|| \boldsymbol \lambda L_i) - \mathbb{E} \left[ KL(Y \,\|\, \boldsymbol \lambda L) \right] \right| \xrightarrow{p} 0,
    \]
    where \( \mathcal{F} \) is the set of feasible nonnegative factorizations.

    \item \textbf{Consistency of $\hat{\boldsymbol \lambda}$.} \\
    By the uniform convergence above and standard M-estimation theory, the minimizers of the empirical KL objective converge in probability to the minimizers of the population KL objective. That is, there exist permutation matrices \( \Pi_N \) and diagonal scaling matrices \( S_N \) that align estimates to the true factorization such that
    \begin{align*}
      \| \hat{\boldsymbol \lambda}S_N \Pi_N - \boldsymbol \lambda^0 \|_F \xrightarrow{p} 0.
    \end{align*}
    This only applies directly to $\hat{\boldsymbol \lambda}$ since it is estimated from the full data matrix and follows from uniform convergence of the empirical loss over $N$ independent observations. Each $\hat L_i$, however, only explicitly depends on the fixed-dimensional vector $Y_i$. We instead treat $\hat L_i$ as a deterministic function of the converging $\hat{\boldsymbol\lambda}$ in the next section. 
    \item \textbf{Consistency of $\hat L_i$.}\\
    The joint minimization implies a marginal convex minimization for each $L_i$ conditional on $\hat{\boldsymbol \lambda}$:
    \begin{align*}
      \hat L_i = \arg \min_{L_i \ge 0} KL(Y_i||\hat{\boldsymbol\lambda} L_i)
    \end{align*}
    Then by stability of convex M-estimators (epiconvergence), $\hat L_i \xrightarrow{p} L_i^0$. Formally:
    \begin{align*}
      &\text{let }M(L_i) = -KL(Y_i||\boldsymbol\lambda^0 L_i), \quad M_n(L_i) = -KL(Y_i||\hat{\boldsymbol\lambda} L_i),\\
      &sup_{L_i \in \mathcal F}\left|M_n(L_i) - M(L_i)\right| \to 0 \text{ because } \hat{\boldsymbol\lambda} \to \boldsymbol\lambda^0,\\
      &L_i^0 = \arg \max_{L_i \ge 0}M(L_i) \text{ is unique, so } M(L_i) < M(L_i^0) \quad\forall L_i \ne L_i^0,\\
      &\text{we define }\hat L_i = \arg \max_{L_i \ge 0}M_n(L_i), \text{ so }M_n(\hat L_i) \ge M_n(L_i^0) - o_P(1).
    \end{align*}
    Which concludes that $\hat L_i \xrightarrow{p} L_i^0$ utilizing Theorem 5.7 of \citet{Vaart_1998}.
    \item \textbf{Convergence of optimization algorithm.} \\
    By assumption, gradient descent converges to a global minimizer of the empirical KL objective. Therefore, the estimates \( (\hat{\boldsymbol \lambda}, \hat{\mathbf L}) \) are consistent up to the equivalence class.    
  \end{enumerate}
  This establishes the consistency of $\hat{\boldsymbol \lambda}, \hat{\mathbf L}$ up to the equivalence class:
  \begin{align*}
    \| \hat{\boldsymbol \lambda} - \boldsymbol \lambda^0 S_N \Pi_N \|_F \xrightarrow{p} 0, \quad \| \hat{\mathbf L} - \Pi_N^{-1} S_N^{-1} \mathbf L^0 \|_F \xrightarrow{p} 0.
  \end{align*}
\end{proof}

\newpage
\section{Simulation studies}

\subsection{Pseudocode: fixing simulation parameters}\label{appendix:sim_params}
Let $K = 5, \mathbf Y =$ somatic mutational counts of 111 breast adenocarcinoma samples, $\boldsymbol \lambda_C =$ COSMIC reference signatures matrix.
\begin{enumerate}
    \item Fit NMF with rank $K$ for a rough view of mutational signatures: $\mathbf Y \approx \hat{\boldsymbol \lambda} \hat{\mathbf L}$
    \item Align estimated signatures to COSMIC reference: 
    \begin{itemize}
        \item Hungarian algorithm on negative cosine similarity matrix between columns of $\hat {\boldsymbol \lambda}$ and $\boldsymbol \lambda_C$ to align signatures
        \item Define $\boldsymbol \lambda$ as a subset of columns of $\boldsymbol \lambda_C$ that align with $\hat{\boldsymbol \lambda}$
    \end{itemize}
    \item Fit NNLS with fixed $\boldsymbol \lambda$: $\mathbf Y  \approx \boldsymbol \lambda \tilde{\mathbf L}$
    \item Define sampling distribution: $p({L})  = \frac{1}{N} \forall L \text{ in columns of } \tilde{\mathbf L}$
\end{enumerate}

\subsection{Pseudocode: simulating a dataset}\label{appendix:sim_data}
The order of $\boldsymbol{\psi}_C$ and $\boldsymbol\Sigma_1$ assumes SBS3 is the last factor. Maximums are taken element-wise.

For $i = 1,\dots, 100$:
\begin{enumerate}
    \item $\ell \sim p(L)$
    \item $L_i \sim \ell + MVN(0, \boldsymbol \Sigma_0), \quad \boldsymbol \Sigma_0 = \sqrt{10}\cdot \mathbf{I}$
    \item ${L}_i(1) \sim \ell + \text{MVN}(\boldsymbol \psi_C, \boldsymbol\Sigma_1),\quad \boldsymbol \psi_C = [0,0,0,0,2000], \quad \boldsymbol \Sigma_1 = \text{diag}(\sqrt{10}, \sqrt{10}, \sqrt{10}, \sqrt{10}, \sqrt{20})$
    \item ${L}_i(0) = \max({L}_i(0), 0),\quad {L}_i(1) = \max({L}_i(1), 0)$
    \item $T_i \sim \text{Bernoulli}(0.2)$
    \item $L_i = L_i(T_i)$
\end{enumerate}
\newpage

\begin{figure}[H]
  \centering
  \includegraphics[width=0.7\linewidth]{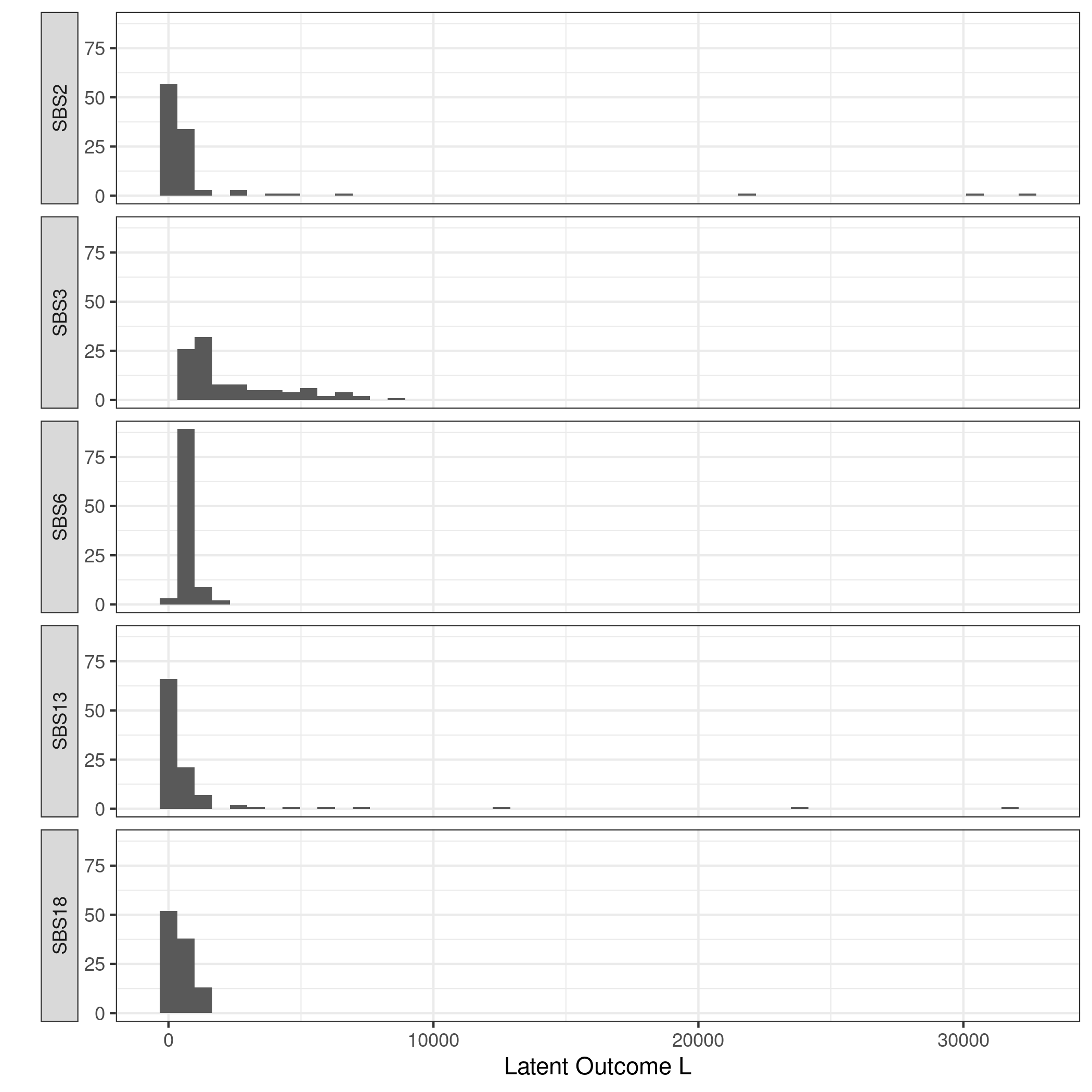}
  \caption{Marginal sampling distributions of each $L_k$, $p(L_k)$. In simulations, the joint distribution $p(L)$ is used to preserve correlation between signature contributions. Outliers in SBS13 and SBS2 are referenced frequently in simulation study results.}
  \label{fig:sim_dist_L}
\end{figure}

\section{Data application}
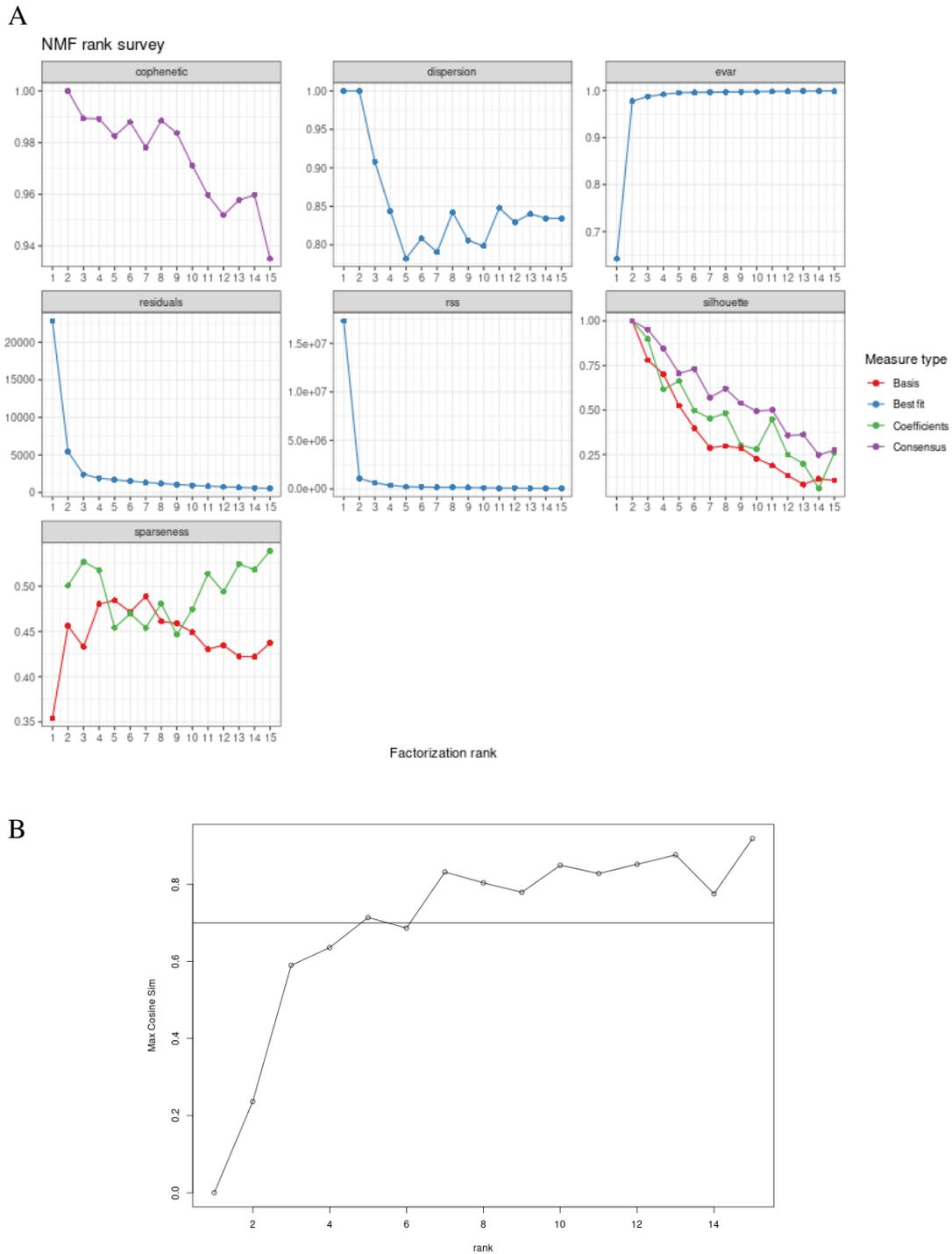
\begin{figure}[H]
  \centering
  \begin{subfigure}{0.9\linewidth}
    \centering
    \begin{picture}(0,0)
      \put(-200,0){\large{A}}
    \end{picture}
    \includegraphics[width=\linewidth,valign=t]{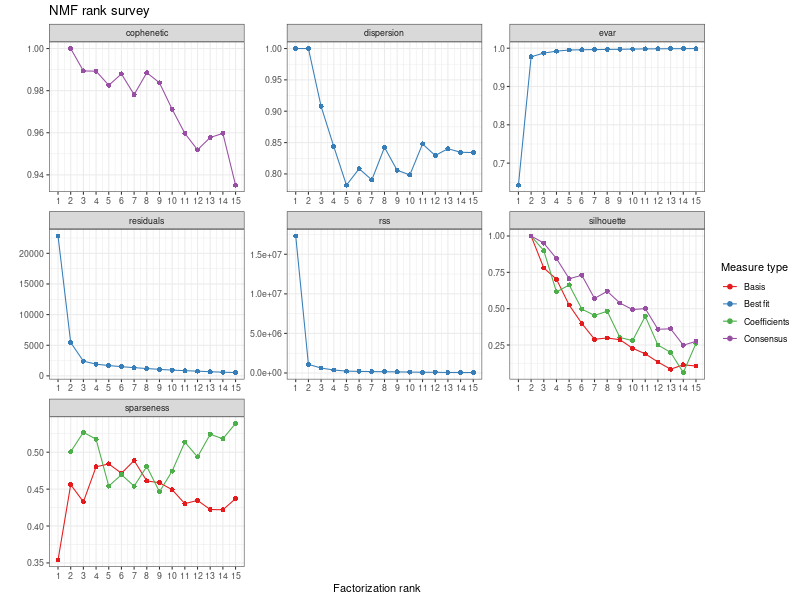}
  \end{subfigure}

  \begin{subfigure}{0.6\linewidth}
    \centering
    \begin{picture}(0,0)
      \put(-200,-30){\large{B}}
    \end{picture}
    \includegraphics[width=\linewidth,valign=t]{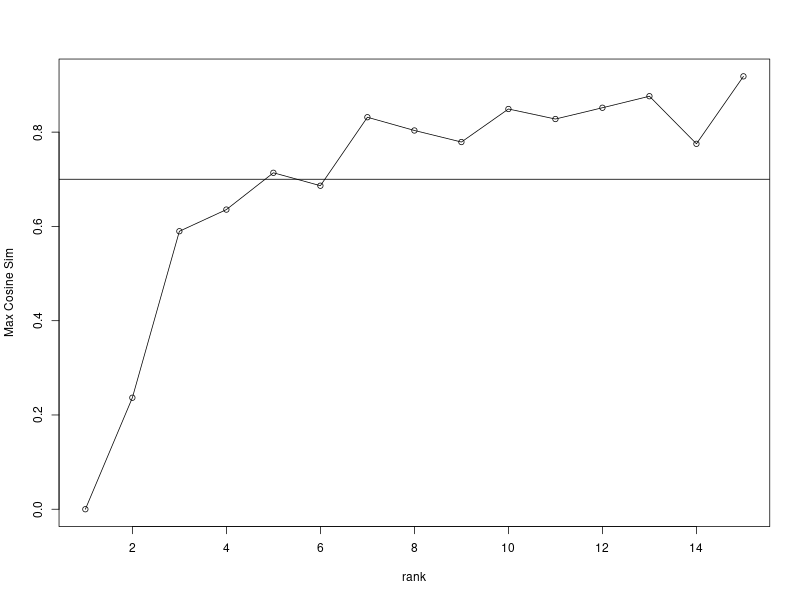}
  \end{subfigure}

  \caption{\textbf{Choosing ranks for breast adenocarcinoma data example}. We determine that ranks between 2 and 5 are reasonable for this dataset to optimize standard metrics while reducing the risk of duplicate signatures. \textbf{A) Standard survey of NMF metrics} for ranks K = 2-15. \textbf{B) Maximum cosine similarity} between estimated signatures for ranks 2-15. High values above 0.7 may indicate duplicate signatures that may hinder interpretability of ATE estimates.}
  \label{fig:rank_sel_nmf_BA_sub45}
\end{figure}

\begin{figure}[H]
  \centering
  \includegraphics[width=0.7\linewidth]{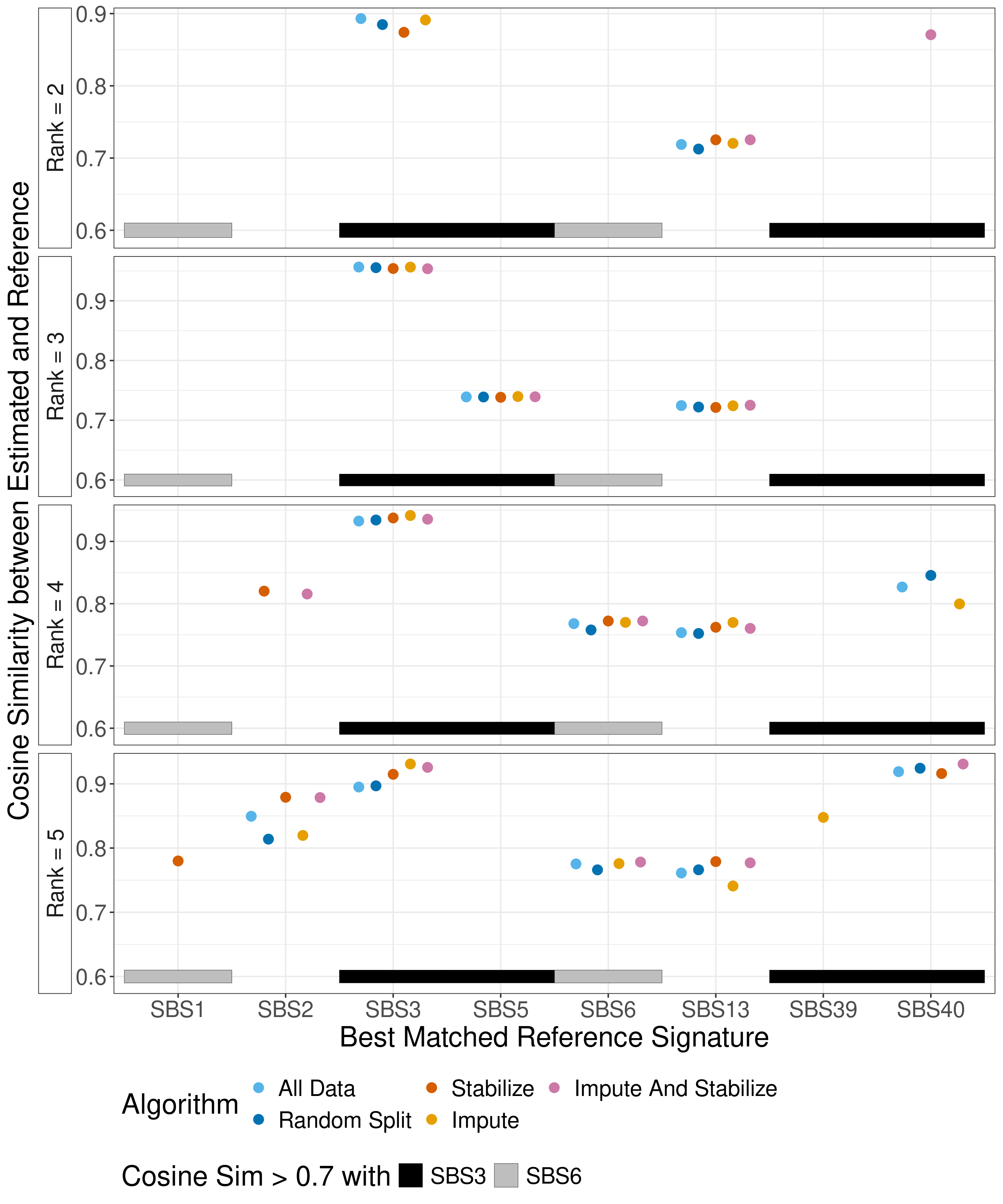}
  \caption{\textbf{Early onset breast adenocarcinoma results}. Cosine similarity between estimated and reference signatures for each rank and each method.}
  \label{fig:application_sim}
\end{figure}

\end{document}